\documentclass[a4paper,11pt]{article}
\usepackage{jheppub}
\usepackage{amsmath,amsthm,amsfonts,amssymb,amscd, mathtools, fancyhdr, color, comment, graphicx, environ, dsfont, relsize}
\usepackage{braket}
\usepackage{cancel,xcolor}
\usepackage[numbers,sort&compress]{natbib}

\title{Log Soft Constraints on KMOC Formalism}
\author[a,b]{Siddhartha Paul,}
\author[a,b]{Adarsh Vishwakarma}

\affiliation[a]{The Institute of Mathematical Sciences
\\IV Cross Road, C.I.T. Campus, Taramani, Chennai 600 113, India}
\affiliation[b]{Homi Bhabha National Institute
\\Training School Complex, Anushakti Nagar, Mumbai 400 094, India}

\emailAdd{siddharthap@imsc.res.in}
\emailAdd{adarshv@imsc.res.in}

\abstract{
The KMOC formalism provides a systematic framework for extracting classical observables perturbatively from on-shell scattering amplitudes. In this work, we apply this formalism to compute electromagnetic observables in four dimensions, focusing in particular on the linear memory effect and its tail contributions. Using the leading and subleading soft-photon theorems to construct the soft radiation kernel, we demonstrate how these infrared observables emerge directly from amplitude data. We further show that demanding the expected non-perturbative properties of memory and tail effects imposes a nontrivial set of consistency conditions on the underlying S-matrix. We interpret these constraints as imposing the requirement of macroscopic causality on the S-matrix via analysis of inclusive observables.
}

\begin{document}

\maketitle
 \date{}
\pagebreak
\newpage

\section{Introduction} \label{introduction}
The recent advances in the understanding and computing scattering amplitudes using on-shell methods have also had rather striking ramifications on our ability to compute classical observables in electromagnetic and gravitational scattering in the Post-Minkowskian (PM) approximation. The building blocks for an amplitude, namely the loop integrands, 
turn into building blocks for classical observables. A carefully defined classical limit at the level of loop integrand, ensures that only a subset of basis integrands survive and hence the increasing sophistication in computing the S-matrix directly aids us in computing classical scattering observables \cite{Neill:2013wsa, Luna:2017dtq, Cheung:2018wkq, Vines:2018gqi, Guevara:2018wpp, Bern:2019crd, Bern:2020uwk, Bern:2020buy, Bern:2020gjj, Bern:2021xze, Bern:2021yeh, Bern:2021dqo, Bern:2022kto, Bern:2023ccb, Bern:2024adl, Damour:2019lcq, Kosower:2018adc, Cristofoli:2021vyo, Arkani-Hamed:2019ymq, Maybee:2019jus, delaCruz:2020bbn, Cristofoli:2021jas, Elkhidir:2023dco, Akhtar:2024mbg, Akhtar:2024lkk, Akhtar:2025fil, Bjerrum-Bohr:2018xdl, Cristofoli:2019neg, Damgaard:2019lfh, Bjerrum-Bohr:2019kec, Cristofoli:2020uzm, Bjerrum-Bohr:2021vuf, Bjerrum-Bohr:2021din, Damgaard:2021ipf, Damgaard:2023ttc, Bini:2024rsy}. In the gravitational theory, amplitude-based techniques can also be employed to analyze the bound orbits in Post-Newtonian (PN) approximation \cite{Porto:2017dgs, Porto:2017shd, Kalin:2019rwq, Kalin:2019inp, Foffa:2019yfl, Kalin:2020mvi}.

Although there are several avenues by which the classical limit of the S matrix can be employed to analyze classical scattering, in this paper, we will focus on one such formalism proposed by Kosower, Maybee, and O'Connell (KMOC). The KMOC formalism is a rather remarkable development in the world of quantum observables, as it defines classical observables as ``inclusive observables" obtained by using the so-called in-in formalism on shell. More in detail, KMOC formulates the computation of classical observables first and foremost as a quantum observable, where for a fixed set of incoming coherent states, one computes the expectation value of any observable in the far future, irrespective of the outgoing state. Classical limit is defined in the large impact parameter approximation, where it can be shown that all the massless momenta become soft as they scale with $\hbar$ \cite{Kosower:2018adc, Cristofoli:2021vyo, Maybee:2019jus, Arkani-Hamed:2019ymq, Monteiro:2020plf, Adamo:2022rmp, DeAngelis:2023lvf, Elkhidir:2023dco, Luna:2023uwd, Brunello:2024ibk, Brunello:2025eso}. 

In a nutshell, KMOC formalism provides us with a map from the space of scattering amplitudes, quantum observables to classical observables such as radiative flux at ${\cal I}^{+}$. The on-shell amplitudes thus replace equations of motion of the classical theory and lead to a catalog of classical observables that can be computed for large impact parameter scattering involving the initial state of two particles. Thus, in the KMOC formalism, classical observables are obtained by taking the $\hbar\, \rightarrow\, 0$ limit of inclusive on-shell observables in the quantum theory, and as a result, we obtain infra-red finite quantities regardless of the fact that the perturbative S-matrix is ill-defined due to infra-red catastrophe.

Classical theory is macro-causal, and at higher orders in perturbation theory, dissipative effects on scattering objects, as well as radiation reaction effects, need to be incorporated in computing the observables. On-shell approach naturally incorporates both of these effects thanks to unitarity and locality of the S-matrix. However, classical observables satisfy macroscopic causality, and it is an interesting question to ask, in what sense does this macroscopic causality emerge from the locality of the quantum S-matrix.

The requirement that classical scattering satisfies macroscopic causality is independent of the perturbative expansion and is encapsulated in the fact that in classical theory, propagation happens in the closure of the future light cone. As the on-shell approach is necessarily perturbative in nature, one may wonder as to how the non-perturbative constraint on classical observables (satisfaction of macroscopic causality) interfaces with the derivation of these observables via the quantum S matrix (which satisfy micro-causality). Our contention in this paper is that one approach to analyzing this question is to demand that radiative observables obtained in the KMOC formalism satisfy classical soft theorems to all orders in the perturbative expansion.

More in detail, in the case of electromagnetic scattering, classical causality and Poincaré Invariance imply that late-time electromagnetic radiation behaves as follows.
\begin{align}
A_{\mu}\, =\, A_{\mu}^{(0)} + \frac{1}{u} A_{\mu}^{(\text{ln})} + \cdots
\end{align}
where the leading order term $A_{\mu}^{(0)}$ in this expansion is known as electromagnetic memory, and the sub-leading term $A_{\mu}^{(\text{ln})}$ is known as log soft factor \cite{Sahoo:2018lxl, Sahoo:2020ryf}.  The actual form of both the memory and the log soft factor is independent of the details of the classical scattering, and both of these observables are completely fixed by the asymptotic momenta and masses of the scattering objects. Clearly, these properties imply that $A_{\mu}^{(0)}, A_{\mu}^{(\text{ln})}$ are completely fixed by macroscopic causality and Poincare invariance. Hence, demanding that the classical limit of the radiative field in the KMOC formalism equal the universal tails to all orders in perturbative expansion can reveal to us how macroscopic causality arises in on-shell approaches. 

In this paper, we initiate an inquiry into this question by subjecting the KMOC formalism to the following criteria: Under what conditions does the classical limit of on-shell amplitude reproduce the electromagnetic memory and the classical log soft photon factor to all orders in perturbative expansion. The first of these constraints was already analyzed in \cite{Bautista:2021llr}, where it was shown that certain moments of the perturbative amplitudes must take a specific form in the classical limit so that electromagnetic memory is reproduced non-perturbatively via on-shell methods. In this paper, we first prove that the local and unitary S-matrix of QED is sufficient to ensure that constraints formulated in \cite{Bautista:2021llr} are satisfied. As a result, the macroscopic causality probed by the existence of the memory observable is, in fact, guaranteed by the micro-causality of the S matrix. However, we then derive the constraints that arise from the universality of the log soft factor. And here, we discover something surprising. Namely, that if on-shell methods are to reproduce the universal log soft factor, then the perturbative amplitudes have to satisfy a set of constraints at all orders in the loop expansion. These constraints are summarized in (\ref{e6.09}). Following \cite{Bautista:2021llr}, we refer to these constraints as sub-leading soft constraints on the KMOC formalism, and as such, they impose macro-causality constraints on the quantum S-matrix.

This paper is organized as follows: in section \ref{s2} we briefly review the KMOC formalism and the aspects of soft theorems which we have used in this paper. Section \ref{s3} describes how the classical limit and the soft limit have been taken to define the soft radiation kernel. In section \ref{s4} we derive the constraints on the leading soft radiation kernel due to the memory term, to all orders in perturbation. Section \ref{s5} and section \ref{s6} describe how the vanishing of the quantum log soft factor puts constraints on the sub-leading soft radiation kernel as well as on the S matrix. Finally, in section \ref{s7} we verify these sub-leading constraints at one loop.

We now summarize the notations used for various quantities throughout the paper. 
\begin{itemize}
\item We define the following measures.
\begin{align}
    d\Phi(p) &= \hat{d}^4 p \, \Theta(p^0) \, \hat{\delta}(p^2-m^2) \nonumber \\
    d\mu(q_i)&=\hat{d}^4q_i\,\hat\delta(2p_i\cdot q_i+q_i^2),\nonumber\\
    d\mu_{q}&=\hat{d}^4q\,\hat\delta(p_1\cdot q)\hat\delta(p_2\cdot q),
\end{align}
where $\hat d^4 q = d^4q/(2\pi)^4$ and $\hat\delta = \delta/2\pi$. Moreover, $d\mu(\bar q_1)$ is defined for scaled $q_i=\hbar\bar q_i.$
\begin{align}
    d\mu(\bar q_i)&=\hat d^4\bar q_i\,\hat\delta(2p_i\cdot \bar q_i+\hbar\bar q_i^2).
\end{align}
    \item We write $L$-loop amplitude compactly as
    \begin{eqnarray}
        \mathcal{A}_{(L)}(P|P') &=& \mathcal{A}_{(L)}(p_1,p_2\rightarrow p_1', p_2').
    \end{eqnarray}
    \item We denote $\mathcal{\tilde A}_{(L)}$ as the amplitude with its coupling stripped off and the most negative power of $\hbar$ factored out, which comes from the ladder diagrams.
\end{itemize}
\section{A brief review on KMOC formalism and soft theorems} \label{s2}
\subsection{KMOC}
In the KMOC formalism \cite{Kosower:2018adc}, classical observables are calculated by taking the classical limit of inclusive on-shell quantum amplitudes. The change in an observable $\hat{O}$ in quantum theory is defined as,
\begin{eqnarray}
	\Delta O  &=& _{\text{in}}\langle \Psi|S^{\dagger}\hat OS|\Psi\rangle_{\text{in}} - \,_{\text{in}}\langle \Psi|\hat{O}|\Psi\rangle_{\text{in}}.\label{e2.1}
\end{eqnarray}
Within this framework, we compute the observables for $2 \rightarrow 2$ scattering in the large impact parameter regime. The particles are in the initial state $ \ket{\Psi}_{\text{in}}$, they evolve to a final state $\ket{\Psi}_{\text{out}}$ by the action of the $S$ matrix. In a frame where the second particle is at the origin, $\ket{\Psi}_{\text{in}}$ is defined as,
\begin{eqnarray}
	|\Psi\rangle_{\text{in}} &=& \int d\Phi(p_1,p_2)\phi_1(p_1)\phi_2(p_2)e^{ib\cdot p_1}\,|p_1,p_2\rangle,\label{e2.2}
\end{eqnarray}
where the measure $d\Phi(p_i)$ imposes on-shell constraints on the momentum eigenstates $\ket{p_1,p_2}$ and the wavepackets $\phi_i(p_i)$. These wavepackets are chosen with appropriate normalization, and they are peaked around the classical momenta of the particles. The width of the wavepackets is regulated by a dimensionless ratio $\xi = l_c/l_w$, where $l_c$ and $l_w$ are respectively the Compton length and position-space wavepacket spread of the particle. Using dimensional analysis, we restore the $\hbar$ factors in the coupling and scale the transfer, off-shell momenta with $\hbar$, essentially writing them in terms of their wavenumber. The $\hbar$ scaling introduces terms that scale as negative powers of $\hbar$, which are the so-called superclassical terms. However, these terms cancel among each other in the classical limit for any physical observables. For the radiation kernel, the cancellation of superclassical terms to all orders in perturbation was shown in \cite{Sinha:2025obs} using $N$-operator formalism. In our calculations, we will ignore these superclassical terms.

The classical limit is taken by imposing the limit $\hbar,\xi \rightarrow 0$. For massive spinning particles, the spin $S$ should also be taken $S \rightarrow \infty$ such that $S\hbar =$  constant \cite{Arkani-Hamed:2019ymq}. We can take $\hat{O}$ to be the momentum operator $\mathds{P}_1^{\mu}$ in (\ref{e2.1}), which gives us the linear impulse $\Delta p^{\mu}_1$ for the first particle. Using $\ket{\Psi}_{in}$ we get the following expression for linear impulse, 
\begin{eqnarray}
	\Delta p_1^{\mu} &=&\, \int  d \mu ( q_1, q_2) \,e^{-ib\cdot   q_1/\hbar} \, \hat\delta^4( q_1+ q_2)\,\Big[i  q^{\mu}_1\,\mathcal{A}(P | P+ Q) \nonumber\\
    &&\,+\,\sum_{X}\!\int\! \prod_{m=1}^Xd\Phi(k_m) \, d\mu( w_1, w_2) \, \hat\delta^4\!\left( w_1+ w_2+\Sigma_{i=1}^X  k_i\right) \nonumber\\
    &&\times\, w_1^{\mu}\mathcal{A}^*(P+ Q | P+ W,\{ k_X\}) \,\mathcal{A}\left(P | P+ W,\{ k_X\}\right)\Big].\label{e2.3}
\end{eqnarray}
We have suppressed the double bracket notation of \cite{Kosower:2018adc} that denotes the wavepacket integration, which in the classical limit, viz. $\xi\rightarrow 0$ amounts to replacing the external momenta with the classical initial momenta. In this paper, we focus on the gauge field at $\mathcal{I^+}$, which is the radiation kernel introduced in \cite{Kosower:2018adc} for the computation of radiation momentum during $2 \rightarrow 3$ scattering and studied in great detail in \cite{Cristofoli:2021vyo}. We substitute the expression for gauge field in terms of creation and annihilation operators in (\ref{e2.1}) and, using (\ref{e2.2}), we write the following,
\begin{eqnarray}
	\mathcal{R}(\bar k)&=&\, \hbar^{3/2} \int d \mu({q}_1,{q}_2)\,e^{-ib\cdot   q_1/\hbar} \ \hat\delta^4( q_1+ q_2+ k) \Big[i\mathcal{A}(P|P+{Q}, k). \nonumber\\
	&&+\,\sum_{X}\!\int\! \prod_{m=1}^Xd\Phi( k_m) d\mu({w}_1,{w}_2) \hat\delta^4\!\left( w_1+ w_2+ k+\Sigma_{i=1}^X k_i\right) \nonumber\\
	&&\times\, \mathcal{A}^*(P+{Q} | P+{W}, \{ k_X\}) \,\mathcal{A}\left(P| P+{W}, k, \{k_X\}\right)\Big].\label{e2.4}
\end{eqnarray}
The gauge field itself is not a physical observable; rather, the radiation flux is what we observe. The flux is on-shell integral over the photon momenta weighted by the factor $|\mathcal{R}(k)|^2$; this imposes the scaling of $\hbar^{-3/2}$ to the kernel.

\subsection{Soft Theorems}\label{s2.2}
Consider a process where $M$ particles interact with each other and produce $N$ particles and an additional photon in the final state. When the photon momentum is taken to be soft compared to the momentum of the hard particles, the amplitude for this process factorizes into a soft factor and a lower-point amplitude. These soft factors capture the radiative effects, and they are defined as the ratio between the $M+N+1$ point and the $M+N$ point amplitudes. These factors have universal and non-universal theory-dependent terms. In $D=4$ as was proved in \cite{Sahoo:2018lxl}, the universal part of the quantum soft photon theorems has the following structure,
\begin{align}
    \mathcal{A}_{M+N+1} (\{p_I\},\{p'_I\},k,\epsilon) &= \sum_{n=-1}^{\infty} \omega^{n}(\ln \omega)^{n+1} \mathcal{S}^{(n)}(\{p_I\},\{p'_I\},k,\epsilon) \mathcal{A}_{M+N}(\{p_I\},\{p'_I\}).
\end{align}
Here, all the momenta are ingoing. The soft factor $\mathcal{S}^{(n)}$ first appears at $(n+1)$-th loop and is $(n+1)$-loop exact. The $\omega^m (\ln \omega)^{m+1}$ factor for $m \le n$ comes after performing the loop integration in the $(n+1)$ loop amplitude. In \cite{Sahoo:2020ryf} and \cite{Sahoo:2018lxl}, the subleading and subsubleading soft factor was shown to be universal and subsequently one-loop and two-loop exact. The subleading soft factor contains a quantum term in addition to the classical term \cite{Sahoo:2018lxl}, whereas the leading soft factor receives no such contribution.
\begin{align}
    \Delta\mathcal{S}^{(-1)}&=0,
\end{align}
\vspace{-7mm}
\begin{align}
    \Delta\mathcal{S}^{(0)}&=\frac{1}{8\pi^2}\sum_{\stackrel{a,b}{b\ne a}}\frac{Q_a^2Q_b}{p_a\cdot k}\ln\left[\frac{p_a\cdot p_b+\sqrt{(p_a\cdot p_b)^2-m_a^2m_b^2}}{p_a\cdot p_b-\sqrt{(p_a\cdot p_b)^2-m_a^2m_b^2}}\right]\frac{m_a^2m_b^2\,\varepsilon_{\mu}k_{\nu}(p_a\wedge p_b)^{\mu\nu}}{((p_a\cdot p_b)^2-m_a^2m_b^2)^{3/2}}.
\end{align}
In \cite{Laddha:2018rle}, the quantum soft factors up to subleading order were shown to be related to the radiative field by defining an appropriate classical limit. In the context of classical mechanics, we can think of a process where charged particles interact with each other. By solving the equation of motion for this process, we can calculate the classical gauge field. This classical gauge field at $\mathcal{I}^{+}$ bears a soft expansion in the far-field limit known as the classical soft photon theorem,
\begin{eqnarray}
    A^{\mu}(k) &=& \sum_{n=-1}^{\infty}\omega^n(\ln\omega)^{n+1}S_{\text{classical}}^{(n),\mu}(\{p_I\},\{p'_I\},k,\epsilon),
\end{eqnarray}
where the leading and subleading terms \cite{Sahoo:2018lxl} are given by
\begin{align}
    \mathcal{S}_{\text{classical}}^{(-1),\mu}&= \sum_{a}Q_a\frac{p_a^{\mu}}{p_a\cdot\hat k},
\end{align}
\vspace{-5mm}
\begin{align}
    \mathcal{S}^{(0),\mu}_{\text{classical}}&=-i\sum_aQ_a\frac{k_{\nu}}{p_a\cdot k}\sum_{\stackrel{b\ne a}{\eta_a\eta_b=-1}}\frac{Q_aQb}{4\pi}\frac{m_a^2m_b^2(p_a\wedge p_b)^{\mu\nu}}{((p_a\cdot p_b)^2-m_a^2m_b^2)^{3/2}}.
\end{align}
The structure of the classical soft factors was conjectured in \cite{Alessio:2024onn} and was proved successively in \cite{Fucito:2024wlg} and \cite{Karan:2025ndk}. We will use these expressions to evaluate the constraints in later sections.
\section{Soft limit in KMOC formalism}\label{s3}

KMOC formalism provides a natural framework to study the connection between classical and quantum soft theorems. Computing the gauge field in KMOC involves taking the classical limit and performing the loop integration. Due to infra-red effects, taking the classical limit before the soft limit is not equivalent to taking them in the reverse order. In our analysis, we take the soft limit prior to loop integration and the classical limit.
\begin{center}\label{f1}
    \begin{figure}[!h]
    \includegraphics[scale=0.30]{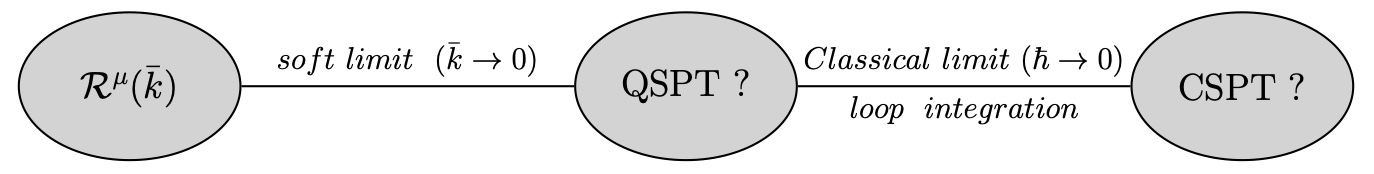}
    \caption{The sequence of classical and soft limits. CSPT = Classical Soft Photon Theorem, QSPT = Quantum Soft Photon Theorem.}
    \end{figure}
\end{center}

The radiation kernel (\ref{e2.4}) in the KMOC formalism has a $5$-point inelastic amplitude in it. We take the soft limit of this $5$-point amplitude at the integrand level. The tree level soft expansion of an $(M+N+1)$-point amplitude is given by,
\begin{align}
    \mathcal{M}_{M+N+1}(\{p_I\},\{p'_I\},k,\epsilon) &= \sum_{n=-1}^{\infty} \omega^n \bigl[\mathcal{\hat S}^{(n)} (\{p_I\},\{p'_I\}, \hat k,\epsilon) \mathcal{M}_{M+N}(\{p_I\},\{p'_I\}) + R_n(\hat k,\epsilon) \bigr].\label{e3.1}
\end{align}
Where $R_n(\hat k, \epsilon)$ has the following form,
\begin{align}
    R_n(\hat k,\epsilon) &= \epsilon^{\mu}\hat k^{\nu_1}\cdots \hat k^{\nu_n}A_{\mu\nu_1 \cdots \nu_{n}}.
\end{align}
Here, $A_{\mu\nu_1\cdots\nu_n}$, anti-symmetric in $\mu$ and $\nu_i$ indices, is a theory-dependent remainder term, which spoils the factorization beyond sub-leading order. The above soft factors capture the universal part of the soft expansion, and have the following form,
\begin{align}
    \mathcal{\hat S}^{(-1)}(\{p_I\},\{p'_I\},k,\epsilon) &= \sum_{a=1}^{M} Q_a \epsilon_{\mu} \frac{p_{a}^{\mu}}{p_a.\hat k} - \sum_{a=1}^{N} Q'_{a} \epsilon_{\mu} \frac{p_{a}'^{\mu}}{p_a'.\hat k},\label{e3.3}
\end{align}
\begin{multline}
    \mathcal{\hat S}^{(n)}(\{p_I\},\{p'_I\},k,\epsilon) = \sum_{a=1}^{M} Q_a \varepsilon_{\mu} \hat k_{\nu} \frac{\hat J^{\mu \nu}_{a}}{p_a.\hat k} \left(\hat k \cdot \frac{\partial}{\partial p_a} \right)^{n} - \sum_{a=1}^{N} Q'_{a} \epsilon_{\mu} \hat k_{\nu} \frac{\hat J'^{\mu \nu}_{a}}{p'_a\cdot \hat k} \left(\hat k \cdot \frac{\partial}{\partial p'_a} \right)^{n}\\ (\forall\, n\ge0).\label{e3.4}
\end{multline}
In (\ref{e3.1}), the amplitudes are unstripped, which take into account the photon momentum in momentum conservation. We use the soft expansion of the amplitude (\ref{e3.1}) up to subleading order in (\ref{e2.4}) to write the soft expansion of the kernel as,
\begin{align}
    \mathcal{R} (k) &= \Big\langle \mathcal{S}^{(-1)} (p_1,p_2,p_1+q_{1},p_2+q_2,k) + \mathcal{\hat S}^{(0)} (p_1,p_2,p_1+q_{1},p_2+q_2,k) \Big \rangle .\label{e3.5}
\end{align}
$\mathcal{\hat S}^{(0)}$ is a differential operator that is linear in the angular momenta of external particles. In the interest of brevity, we will henceforth simply use $\langle \rangle$ to denote the classical limit, 
\begin{align*}
    \Big \langle \hat f \Big \rangle &= \int d \mu ( q_1, q_2) \,e^{-ib\cdot   q_1/\hbar} \, \Big[i \hat f(q_1,q_2) \mathcal{M}(P | P+ Q) \nonumber\\
    &\,+\,\sum_{X}\int\! \prod_{m=1}^Xd\Phi(k_m) \, d\mu( w_1, w_2) \nonumber\\
    &\times\,\mathcal{M}^*(P+ Q | P+ W,\{ k_X\}) \,\hat f(w_1,w_2) \mathcal{M}\left(P | P+ W,\{ k_X\}\right)\Big].
\end{align*}
In later sections, we will suppress the $\hbar\rightarrow0$ limit and will make it explicit only while specifying certain results. In classical theory, the contribution to the soft expansion (\ref{e3.5}) of the gauge field comes from the matter poles in the loop integrals. The contributions from photon poles cancel among each other in the perturbative expansion, as in the classical theory, the photon propagators are retarded. Each of the soft factors is built out of a certain combination of transfer momenta and angular momenta. This gives us two classes of constraints: one, due to the vanishing of terms coming from photon poles in the classical limit, and second, due to the fact that the contribution from matter poles matches the classical expression. 
\section{Leading Soft Constraints on $\mathcal{R}(k)$}\label{s4}

In \cite{Bautista:2021llr}, it was shown that the leading classical soft photon theorem imposes a hierarchy of constraints on the conservative sector in KMOC formalism. More in detail, it was argued that the universal form of electro-magnetic memory for a fixed set of initial and final configurations of charged particles leads to the following set of constraints on a set of observables.
\begin{align}
     \Big \langle q^{\mu_1}_a\cdots q_a^{\mu_N} \Big \rangle &= \Delta p^{\mu_1}_a \cdots\Delta p^{\mu_N}_a, \label{e4.1a}
\end{align}
where $\Big \langle q^{\mu_1}_a\cdots q_a^{\mu_N} \Big \rangle$ is defined as in (\ref{e4.5}) These constraints were verified to next-to-leading order (NLO) in \cite{Bautista:2021llr}. In this section, we give an explicit proof for (\ref{e4.1a}) to all orders in perturbation theory.\\
The leading soft factor for $2\, \rightarrow\, 2$ scattering (\ref{e3.4}) where the final momenta are given by $p'_a=p_a+q_a$ can be written as,
\begin{align}
    \mathcal{S}_a^{(-1)}(p_1,p_2,p_1+q_{1},p_2+q_2,k)&=Q_a \epsilon_{\mu} \biggl[-\frac{q_{a}^{\mu}}{p_a\cdot k} + \sum_{N=1}^{\infty}(-1)^{N+1} \frac{(q_a\cdot k)^N}{(p_a\cdot k)^{N+1}}(p_a^{\mu} + q_{a}^{\mu})\biggr].
\end{align}
We write the leading-order contribution to the kernel for particle $a$ as,
\begin{eqnarray}
\mathcal{R}_a^{(-1)}(k) &=& \Big \langle \mathcal{S}_a^{(-1)}(p_1,p_2,p_1+\hbar \bar{q}_1,p_2+\hbar \bar{q}_2,k) \Big \rangle.\label{e4.2}
\end{eqnarray}
The above kernel (\ref{e4.2}) equals the classical LO soft factor, giving us the leading soft constraint. 
\begin{eqnarray}
\mathcal{R}_a^{(-1)}(k) &=& \mathcal{S}_a^{(-1)}(p_1,p_2,p_1+\Delta p_1,p_2+\Delta p_2,k) .\label{e4.3}
\end{eqnarray}
To show that (\ref{e4.2}) reduces to (\ref{e4.3}), first we prove that for any rational function $F(q_1,q_2 = - q_{1} - k)$, we will have $\Big\langle F(q_1,q_2)\Big\rangle=F(\Delta p_1,\Delta p_2)$. In order to calculate this, we need the expectation value of the $n$-th moment of the transfer momentum $\Big \langle q^{\mu_1}_a\cdots q_a^{\mu_N} \Big \rangle$. We get the moment of degree $1$ i.e. $\Big \langle q_a^{\mu} \Big \rangle$ by taking the expectation value of the operator $S^{\dag}[\mathds{P}_a^{\mu},S]$ over the two particle initial state $\ket{\Psi}_{\text{in}}$. Similarly, it is straightforward to see that the $n$-th moment arises when we take the expectation value of the operator $S^{\dag}[\mathds{P}_a^{\mu_1},\ldots ,[\mathds{P}_a^{\mu_N},S]\ldots]$.
\begin{align}
     \Big \langle q^{\mu_1}_a\cdots q_a^{\mu_N} \Big \rangle &= \;\; \lim_{\hbar \rightarrow 0} \bra{\Psi_{\text{in}}} S^{\dag}[\mathds{P}_a^{\mu_1},\ldots,[\mathds{P}_a^{\mu_N},S]\ldots]\ket{\Psi_{\text{in}}} .
    \label{e4.4}
\end{align}
In terms of the phase-space integral, we can write (\ref{e4.4}) as follows.
\begin{align}
    \Big \langle q^{\mu_1}_a\cdots q_a^{\mu_N} \Big \rangle &=  \int d\mu(q_1,q_2) \,\hat\delta^4(q_1+q_2)e^{-iq_1\cdot b/\hbar}\Biggr[\,\,i q^{\mu_1}_a\cdots q_a^{\mu_N}  \mathcal{A}(P| P+Q) \nonumber \\
    &+ \sum_{X}\int \prod_{m=0}^{X} d\Phi(k_m) \, d\mu(w_1,w_2) \,\hat\delta^4(w_1+w_2+\Sigma_m \,k_m)\,\,w_a^{\mu_1}\cdots w_a^{\mu_N}\nonumber \\
    & \times \mathcal{A}(P | P+W,\{k_X\})  \mathcal{A}^*(P+Q|P+W,\{k_X\})\Biggr].\label{e4.5}
\end{align}
The expectation value in (\ref{e4.4}) over the states $|\Psi_{\text{in}}\rangle$ reduces to the product of impulses up to an additive quantum factor. The quantum factor depends on the remainder terms $\mathfrak{R}^{\mu_1 \cdots \mu_k}_a \;\; \forall \;\; k\geq2$, which we define as follows,
\begin{align}
    \mathfrak{R}^{\mu_1 \cdots \mu_k}_a &= \sum_{\alpha_1 \cdots \alpha_{n-1} \neq \ket{\Psi_{\text{in}}}} \bra{\Psi_{\text{in}}}S^{\dag} \mathds{P}^{\mu_1}_a S\ket{\alpha_1}\bra{\alpha_1} S^{\dag} \mathds{P}^{\mu_2}_a S\ket{\alpha_2} \cdots \bra{\alpha_{k-1}}S^{\dag}\mathds{P}^{\mu_k}_a S\ket{\Psi_{\text{in}}}. \label{e4.6}
\end{align}
The states $\ket{\alpha_i}$ are the (over)complete set of coherent states. $\mathfrak{R}^{\mu_{1}\, \cdots\, \mu_{k}}_a$ is related to the classical observable, $\Big \langle q^{\mu_1}_a \cdots q_a^{\mu_N} \Big \rangle$ through $f^{\mu_1 \cdots \mu_N}_a$,
\begin{align}
    f^{\mu_1 \cdots \mu_N}_a &= \sum_{\sigma \in S_N} C_1(\sigma) \ \Delta p^{\mu_{\sigma(1)}}_a \cdots \Delta p^{\mu_{\sigma(N-2)}}_a \mathfrak{R}^{\mu_{\sigma(N-1)} \mu_{\sigma(N)}}_a \nonumber \\
    & + \sum_{\sigma \in S_N} C_2(\sigma) \ \Delta p_a^{\mu_{\sigma(1)}} \cdots \Delta p_a^{\mu_{\sigma(N-3)}} \mathfrak{R}^{\mu_{\sigma(N-2)} \ \mu_{\sigma(N-1)} \ \mu_{\sigma(N)}}_a \nonumber \\ 
    &+ \sum_{\sigma \in S_N} C_3(\sigma) \ \Delta p_a^{\mu_{\sigma(1)}} \cdots p_a^{\mu_{\sigma(N-3)}} \mathfrak{R}^{\mu_{\sigma(N-2)} \ \mu_{\sigma(N-1)} \ \mu_{\sigma(N)}}_a + \cdots + \mathfrak{R}^{\mu_{1} \cdots \mu_{N}}_a. \label{e4.8}
 \end{align}
We can separate $f_a^{\mu_1 \cdots \mu_N}$ from the expectation value of the nested commutator in (\ref{e4.4}) by expanding the commutator brackets and then inserting over-complete set of coherent states in between them. We then write the coherent states in terms of $\ket{\Psi_{\text{in}}}$ and states that are orthogonal to it. The expectation values over the in-states separate to give the product of linear impulses, and we have the following decomposition in the classical limit,
\begin{align}
     \Big \langle q^{\mu_1}_a\cdots q_a^{\mu_N} \Big \rangle &= \Delta p^{\mu_1}_a \cdots\Delta p^{\mu_N}_a + \lim_{\hbar \rightarrow 0} f_a^{\mu_1 \cdots \mu_N}. \label{e4.9}
\end{align}
It is straightforward to generalize the above result for moments containing $q_1$ and $q_2$. The expectation $\Big \langle F(q_1,q_2) \Big \rangle$ is essentially the sum over all the moments. Thus, we have $\Big \langle F(q_1,q_2) \Big \rangle = F(\Delta p_1,\Delta p_2)$. From the above result, it follows that (\ref{e4.2}) in the classical limit reduces to (\ref{e4.3}). In the following subsection, we verify that the remainder term $\mathfrak{R}^{\mu_1\mu_2}_a$ is quantum.

\subsection{$\Big \langle q^{\mu_1}_a\cdots q^{\mu_N}_a \Big \rangle $ and Contact terms}

For the operator $S^{\dag}[\mathds{P}^{\mu_1}_a,[\mathds{P}^{\mu_2}_a,S]]$, we use the over-complete set of states to write the following,
\begin{align}
    \bra{\Psi_{\text{in}}} S^{\dag}[\mathds{P}^{\mu_1}_a,[\mathds{P}^{\mu_2}_a,S]]\ket{\Psi_{\text{in}}} &= \Delta p^{\mu_1}_a \Delta p^{\mu_2}_a + \mathfrak{R}^{\mu_1 \mu_2}_a, \label{e4.10}
\end{align}
where the remainder term is given by
\begin{align}
    \mathfrak{R}^{\mu_1 \mu_2}_a = \sum_{\alpha \neq \Psi_{\text{in}}} \bra{\Psi_{\text{in}}}S^{\dag} \mathds{P}^{\mu_1}_a S\ket{\alpha}\bra{\alpha} S^{\dag} \mathds{P}^{\mu_2}_a S\ket{\Psi_{\text{in}}}.\label{e4.11}
\end{align}
The state $\ket{\alpha}$ can be expressed as a sum over the states in which the momenta of the matter particles peak around their classical value, and the states for which the wavepackets of the matter particles have negligible overlap with $\phi_{\text{in}}$.
\begin{align*}
    \ket{\alpha} &= \sum_{X} \int \prod_{i=1}^{2} d\Phi(p_i) \,\, \phi_{in}(p_i) e^{-i \frac{p_1.b}{\hbar}} \ket{p_1\,p_2\,X} + \sum_{Y} \int \prod_{i=1}^{2} d\Phi(r_i) \,\, \psi_{i}(r_i) e^{-i \frac{r_1.b}{\hbar}} \ket{r_1\,r_2\,Y}.
\end{align*}
The above expression with no intermediate photons ($X=0$) produces the product of impulses in (\ref{e4.9}). The negligible overlap of $\psi_i$ and $\phi_{\text{in}}$ translates to the following no-overlap condition in the classical limit,
\begin{align}
    \lim_{\hbar \rightarrow 0} \int d \Phi(p) \, \phi_{\text{in}}^*(p) \, \psi_i(p') &= 0.\label{e4.12}
\end{align}
This ensures that these two wavepackets are not peaked at the same momentum value, and they don't contribute classically. Now using the expression for $\ket{\alpha}$ we evaluate (\ref{e4.1a}),
\begin{align}
    \mathfrak{R}^{\mu_1 \mu_2}_a &= \sum_{X,Y} \int \prod_{j=1}^{2} d\Phi(\tilde{p}_j) d\Phi(\tilde{p}'_j) d\Phi(p_j) d\Phi(p'_j) \, \phi_{in}(p_j)\phi^{*}_{in}(p'_j) \phi^{*}_{in}(\tilde{p}_j) \phi_{in}(\tilde{p}'_j) \nonumber \\
    & \qquad \times e^{-i\frac{(p_1-p'_1)b}{\hbar}}  \bra{\tilde{p}_1\,\tilde{p}_2}S^{\dag}\mathds{P}^{\mu_1}_a S\ket{p_1\,p_2\,X} \bra{p'_1\,p'_2\,Y}S^{\dag}\mathds{P}^{\mu_2}_a S\ket{\tilde{p}'_1\,\tilde{p}'_2} \nonumber \\
    & + \sum_{X,Y} \int \prod_{j=1}^{2} d\Phi(\tilde{p}_j) d\Phi(\tilde{p}'_j) d\Phi(p_j) d\Phi(r'_j) \, \phi^{*}_{in}(\tilde{p}_j) \phi_{in}(\tilde{p}'_j) \phi_{in}(p_j) \psi^*(r'_j) \nonumber \\
    & \qquad \times e^{-i\frac{(p_1-r'_1)b}{\hbar}} \bra{\tilde{p}_1\,\tilde{p}_2} S^{\dag} \mathds{P}^{\mu_1}_a S \ket{p_1\,p_2\,X} \bra{r'_1\,r'_2\,Y} S^{\dag} \mathds{P}^{\mu_2}_a S \ket{\tilde{p}'_1\,\tilde{p}'_2} \nonumber \\
    & + \sum_{X,Y} \int \prod_{j=1}^{2} d\Phi(\tilde{p}_j) d\Phi(\tilde{p}'_j) d\Phi(p'_j) d\Phi(r_j) \, \phi^{*}_{in}(\tilde{p}_j) \phi_{in}(\tilde{p}'_j) \phi^*_{in}(p'_j) \psi(r_j) \nonumber \\
    & \qquad \times e^{-i\frac{(p'_1-r_1)b}{\hbar}}  \bra{\tilde{p}_1\,\tilde{p}_2} S^{\dag} \mathds{P}^{\mu_1}_a S \ket{r_1\,r_2\,Y} \bra{p'_1\,p'_2\,X} S^{\dag} \mathds{P}^{\mu_2}_a S \ket{\tilde{p}'_1\,\tilde{p}'_2} \nonumber \\
    & + \sum_{X,Y} \int \prod_{j=1}^{2} d\Phi(\tilde{p}_j) d\Phi(\tilde{p}'_j) d\Phi(r_j) d\Phi(r'_j) \, \phi^{*}_{in}(\tilde{p}_j) \phi_{in}(\tilde{p}'_j) \psi(r_j) \psi^*(r'_j) \nonumber \\
    & \qquad \times e^{-i\frac{(r_1-r'_1)b}{\hbar}} \bra{\tilde{p}_1\,\tilde{p}_2} S^{\dag} \mathds{P}^{\mu_1}_a S \ket{r_1\,r_2\,X} \bra{r'_1\,r'_2\,Y} S^{\dag} \mathds{P}^{\mu_2}_a S \ket{\tilde{p}'_1\,\tilde{p}'_2}.\label{a.4}
\end{align}
In the classical limit, the last three terms vanish due to the no-overlap condition (\ref{e4.12}). The first term is quantum when we write $p_a',\tilde{p}_a'$ as $p_a + \mathcal{O}(\hbar)$, the measures $d \Phi(p_a')$ and $d \Phi(\tilde{p}_a')$ will contribute an additional $\hbar^3$ factor each and for $\bra{\tilde{p}_1\,\tilde{p}_2} S^{\dag} \mathds{P}^{\mu_1}_a S \ket{p_1\,p_2\,X}$ we have the following expression,
\begin{align*}
    \bra{\tilde{p}_1\,\tilde{p}_2} S^{\dag} \mathds{P}^{\mu_1}_a S \ket{p_1\,p_2\,X} &= p^{\mu_1}_a \braket{\tilde{p}_1\,\tilde{p}_2|p_1\,p_2\,X} + i(\tilde{p}^{\mu_1}_a - p^{\mu_1}_a) \braket{\tilde{p}_1\,\tilde{p}_2|T|p_1\,p_2\,X} \\
    & + \sum_W \int \prod_{j=1}^{2} d\Phi(s_j) (s_a^{\mu_1}-p_a^{\mu_1}) \braket{\tilde{p}_1\,\tilde{p}_2|T^{\dag}|s_1\,s_2\,W} \braket{s_1\,s_2\,W|T|p_1\,p_2\,X}.
\end{align*}
$s_a^{\mu}$ are the momenta of the matter particles present in the intermediate states. When we scale these momenta with $\hbar$, the leading term in $\bra{\tilde{p}_1\,\tilde{p}_2} S^{\dag} \mathds{P}^{\mu_1}_a S \ket{p_1\,p_2\,X}$ becomes of $\mathcal{O}(\hbar^{-3})$ which is not sufficient to compensate the $\hbar$ factors coming from the measures. Hence
\begin{align}
    \lim_{\hbar\rightarrow0} \,\,\mathfrak{R}^{\mu_1\mu_2}_a = 0.\label{a.5}
\end{align}
Similarly, for $S^{\dag} [\mathds{P}^{\mu_1}_a , [\mathds{P}^{\mu_2}_a , [\mathds{P}^{\mu_3}_a , S]]]$ and $S^{\dag} [\mathds{P}^{\mu_1}_a, [\mathds{P}^{\mu_2}_a, [\mathds{P}^{\mu_3}_a [\mathds{P}^{\mu_4}_a, S ]]]]$, the remainder terms take the following form
\begin{align*}
    \bra{\Psi_{\text{in}}} S^{\dag} [\mathds{P}^{\mu_1}_a , [\mathds{P}^{\mu_2}_a , [\mathds{P}^{\mu_3}_a , S]]]\ket{\Psi_{\text{in}}} &= \Delta p^{\mu_1}_a \Delta p^{\mu_2}_a \Delta p^{\mu_3}_a + \Delta p^{\mu_3}_a \mathfrak{R}^{\mu_1 \mu_2}_a + \Delta p^{\mu_1}_a \mathfrak{R}^{\mu_2 \mu_3}_a \\
    &- p^{\mu_2}_a \mathfrak{R}^{\mu_1 \mu_3}_a + \mathfrak{R}^{\mu_1 \mu_2 \mu_3}_a, \\
    \bra{\Psi_{\text{in}}} S^{\dag} [\mathds{P}^{\mu_1}_a, [\mathds{P}^{\mu_2}_a, [\mathds{P}^{\mu_3}_a [\mathds{P}^{\mu_4}_a, S ]]]] \ket{\Psi_{\text{in}}} &= \Delta p^{\mu_1}_a \Delta p^{\mu_2}_a \Delta p^{\mu_3}_a \Delta p^{\mu_4}_a + \Delta p^{\mu_1}_a \Delta p^{\mu_2}_a \mathfrak{R}^{\mu_3 \mu_4}_a \\
    &+ \Delta p^{\mu_1}_a \Delta p^{\mu_4}_a \mathfrak{R}^{\mu_2 \mu_3}_a - \Delta p^{\mu_1}_a p^{\mu_3}_a \mathfrak{R}^{\mu_2 \mu_4}_a \\
    &- p^{\mu_2}_a \Delta p^{\mu_4}_a \mathfrak{R}^{\mu_1 \mu_3}_a + p^{\mu_2}_a p^{\mu_3}_a \mathfrak{R}^{\mu_1 \mu_4}_a + \Delta p^{\mu_1}_a \mathfrak{R}^{\mu_2 \mu_3 \mu_4}_a \\ 
    &+ \Delta p^{\mu_4}_a \mathfrak{R}^{\mu_1 \mu_2 \mu_3}_a - p^{\mu_3}_a \mathfrak{R}^{\mu_1 \mu_2 \mu_4}_a - p^{\mu_2}_a \mathfrak{R}^{\mu_1 \mu_3 \mu_4}_a\\
    &+ \mathfrak{R}^{\mu_1\mu_2}_a \mathfrak{R}^{\mu_3\mu_4}_a + \mathfrak{R}^{\mu_1\mu_2\mu_3\mu_4}_a.
\end{align*}
Here, the extra terms in the above expressions reduce to $f_a^{\mu_1\mu_2\mu_3}$ and $f_a^{\mu_1\mu_2\mu_3\mu_4}$ using (\ref{e4.8}). Similarly the expression for $\mathfrak{R}^{\mu_1 \cdots \mu_N}_a$ will contain one quantum term, and the remaining $(4^{N-1}-1)$ terms will vanish due to the no-overlap condition.
\begin{align}
    \lim_{\hbar \rightarrow 0} \mathfrak{R}^{\mu_1...\mu_N}_a = 0.\label{e4.15}
\end{align}
\section{Sub-leading soft radiation kernel} \label{s5}
As reviewed in subsection (\ref{s2.2}), classical soft photon theorems in $D = 4$ are an infinite hierarchy of universal factorization theorems which state that in a soft expansion of the radiative field, coefficients of $\omega^{n}(\ln\omega)^{n+1}$ are universal $\forall\, n \geq\, -1$. In this section, we subject the radiation kernel obtained using the KMOC formalism to the following consistency criteria: To all orders in perturbative expansion, the sub-leading term in the kernel should precisely match the so-called log soft radiative mode. As we show below, this criterion leads to a set of non-trivial constraints that the hard amplitude must satisfy at any order in perturbative expansion. Before going into the details of the derivation, we outline the strategy used in deriving these consistency conditions. 

\begin{itemize}
\item Our approach to the soft limit of the radiation kernel starts with the sub-leading term in the soft expansion of the inelastic loop integrand. That is, we start by expanding the tree-level sub-leading soft factor in the limit of small exchange momenta. Thus, for the first particle, the soft factor can be written as, 
    \begin{align}
        \mathcal{\hat S}^{(0)}_{1} (p_1,p_1+q_{1},k) &= Q_1 \varepsilon_{\mu}k_{\nu} \biggl[-\frac{\Delta \hat J^{\mu \nu}_1}{p_1\cdot k} + \sum_{N=1}^{\infty}(-1)^{N+1} \frac{(q_1\cdot k)^N}{(p_1\cdot k)^{N+1}} \hat J^{\mu \nu}_{1,\text{out}}\biggr].\label{e5.1}
    \end{align}
    With the substitution $p'_1=p_1+q_1$, the angular impulse and the final angular momentum are written as \cite{Akhtar:2024mbg},
    \begin{align}
        \Delta \hat J^{\mu\nu}_1 &= -i\hbar \bigl(p_1 \wedge \partial_{p_1} + q_1 \wedge \partial_{q_1}\bigr)^{\mu \nu},\label{e5.2} \\
        \hat J^{\mu \nu}_{1,\text{out}}&= -i\hbar\bigl((p_1+q_1) \wedge \partial_{q_1}\bigr)^{\mu \nu}.\label{r5.3}
    \end{align}
    \item The subleading soft factor can be isolated from the radiation kernel (\ref{e3.5}) using the operator $\lim_{\omega\rightarrow0}\,\omega(\partial_{\omega})^2\omega$. 
    \begin{align}
        \mathcal{R}^{(0)} (k) &=\lim_{\omega\rightarrow0}\,\omega(\partial_{\omega})^2(\omega\mathcal{R}(k)).
    \end{align}
    The subleading kernel, as can be seen in (\ref{e5.1}) gets contribution from two quantities $\langle \Delta \hat J^{\mu\nu}_1\rangle$ and $\langle q_1^{\mu_1}\cdots q_1^{\mu_N}\hat J^{\mu\nu}_{1,\mathrm{out}}\rangle$.
    \begin{align}
        \mathcal{R}^{(0)} (k) &= Q_1 \varepsilon_{\mu}k_{\nu} \biggl[-\frac{\Big \langle \Delta \hat J^{\mu \nu}_1 \Big \rangle}{p_1\cdot k} + \sum_{N=1}^{\infty}(-1)^{N+1} \frac{k_{\alpha_1} \cdots k_{\alpha_N}}{(p_1\cdot k)^{N+1}} \Big \langle q_1^{\alpha_1}\cdots q_1^{\alpha_N}\hat J_{1,\text{out}}^{\mu \nu} \Big \rangle \biggr].\label{e5.4}
    \end{align}
    Here we have suppressed the particle label.
    \item  We show that the logarithmic contribution to the radiation kernel arises from the region of integration for loop momenta to $\omega<|l|<<|q|$.
    \item We finally show that (\ref{e5.4}) leads to the classical log soft factor if and only if the $2 \rightarrow\, 2$ scattering amplitude satisfies the following set of constraints, 
\begin{align}\label{e5.5}
{\cal Z}_{i}(p_{1}, p_{2}, b\,; L)\, &=\, 0,\; \forall\, 1\, \leq\, L\;\;(i=1,2,3),\nonumber\\
\Big \langle q_1^{\alpha_1}\cdots q_1^{\alpha_N}\hat J_{1,\text{out}}^{\mu \nu} \Big \rangle_{(L),\,\text{q}}&=0,\nonumber\\
{\cal R}^{(0)}_{\ln}\vert_{{\mathcal{Z}_i} = 0}\, &=\, \textrm{Classical log soft factor}.
\end{align}
Detailed expressions for $\mathcal{Z}_i$ can be found in section (\ref{s6}).

\item In section (\ref{s7}), we show that any amplitude that satisfies the first set of vanishing conditions in (\ref{e5.5}) also satisfies the second condition at one loop, thus showing that the KMOC formalism leads to the classical log soft factor.  
\end{itemize}
\subsection{Computation of Angular Impulse}

In this section, we compute the angular impulse $\langle\Delta \hat J^{\mu\nu}_1\rangle = R^{\mu\nu}+C^{\mu\nu}$. It can be easily checked that $\langle\Delta\hat J^{\mu\nu}_1\rangle$ can be expressed in terms of the amplitude as follows.
\begin{align}
    \langle\Delta\hat J_1^{\mu \nu}\rangle &= \hbar\int d\mu(q_1,q_2) \ e^{-i b \cdot q_1/\hbar} \Big[(p_1\wedge \partial_{p_1}+q_1\wedge \partial_{q_1})^{\mu\nu} \bigl\{\delta^4(q_1+q_2)\, \mathcal{A}(P|P+Q) \bigr\}\nonumber\\
    &-i\sum_{X}\int d\mu(w_1,w_2) \ \delta^4(q_1+q_2-w_1-w_2 -\{k_X\})\,\mathcal{A}^*\bigl(P+W,\{k_X\} | P+Q\bigr) \nonumber \\ 
    & \times (p_1\wedge \partial_{p_1}+w_1\wedge \partial_{w_1})^{\mu\nu} \bigl \{ \delta^4(w_1+w_2+\{k_X\}) \ \mathcal{A}\bigl(P|P+W,\{k_X\}\bigr) \bigr\}\Big] .\label{e5.6}
\end{align}

Here, $R^{\mu \nu}$ and $C^{\mu \nu}$ are the linear and cut contributions respectively. The $L$th loop contribution to $R^{\mu \nu}$ can be written in the following form
\begin{align}
	\Aboxed{R_{(L)}^{\mu\nu}&=\frac{e^{2L+2}}{4\hbar^{L}}\int d\mu_{\bar q}\,e^{-ib\cdot\bar q}\sum_{n=0}^L\frac{(-1)^n\,\hbar^n}{2^n\,n!}\,\left(F_{1,n}^{\mu\nu(L)}-F_{2,n}^{\mu\nu(L)}\right),\hspace{2mm}(L\ge0).}\label{e5.7}
\end{align}

The cut contribution $C^{\mu \nu}_{(L)}$ can also be written in the same form as
\begin{align}
	\Aboxed{C_{(L)}^{\mu\nu} &=\frac{e^{2L+2}}{4\hbar^{L}}\int d\mu_{\bar q}\,e^{-i b\cdot \bar q}\sum_{n=0}^L\frac{(-1)^n\,\hbar^n}{2^n\,n!}\Big(G_{1,n}^{\mu\nu(L)}-G_{2,n}^{\mu\nu(L)}\Big),\hspace{2mm}(L\ge1).}\label{e5.8}
\end{align}

The expressions for $F_{i,n}^{(L)}$ and $G_{i,n}^{(L)}$ are given in the Appendix (\ref{A}). Several comments are in order.
\begin{itemize}
\item To write $R^{\mu \nu}_{(L)}$ and $C^{\mu \nu}_{(L)}$ in the current form, we have used IBP to transfer derivatives from the momentum-conserving delta functions. Moreover, the derivatives $\partial_{x_i}$ are taken after decomposing $q$ as 
\begin{eqnarray}
q^{\mu}&=&\alpha_1 p_1^{\mu}+\alpha_2 p_2^{\mu}+q_{\perp}^{\mu}.\label{e5.9}
\end{eqnarray}
The coefficients $\alpha_i$ are given by 
\begin{eqnarray}
    \alpha_1&=&\frac{1}{D}\left((p_1\cdot p_2)\,x_2-m_2^2x_1\right),\\
    \alpha_2&=&\frac{1}{D}\left((p_1\cdot p_2)\,x_1-m_1^2x_2\right).
\end{eqnarray}
where $D=(p_1\cdot p_2)^2-m_1^2m_2^2$ \cite{Bautista:2021llr}. 
\item To calculate, e.g. $R_{(L)}$, we expand the amplitude up to $\mathcal{O}(\hbar^L)$. The terms of $\mathcal{O}(\hbar^{<L})$ will give rise to superclassical terms in addition to the classical terms when multiplied with appropriate powers of $\hbar$. The vanishing of superclassical terms for the radiation kernel was shown in \cite{Sinha:2025obs} using $N$-operator formalism. The terms containing $(\bar{q} \wedge b)^{\mu \nu}$ and $(\bar{w} \wedge b)^{\mu \nu}$ in the linear and cut terms respectively combine to give $(\Delta p \wedge b)^{\mu \nu}$, which is $\mathcal{O}(\omega^0)$ and do not contribute at $O(\log \omega)$.
\end{itemize}
\subsection{Computation of $\langle q_1^{\alpha_1}\cdots q_1^{\alpha_N}\hat J_{1,\text{out}}^{\mu \nu} \rangle$.} 
The second term in (\ref{e5.4}) has the following expression in terms of amplitude.
\begin{align}
    \Big\langle q_1^{\alpha_1}\cdots&\, q_1^{\alpha_N}\hat J_{1,\text{out}}^{\mu \nu} \Big\rangle \nonumber\\
    &= \hbar\int d\mu(q_1,q_2) \ e^{-i b \cdot q_1/\hbar} \Big[q_1^{\alpha_1}\cdots q_1^{\alpha_N}((p_1+q_1)\wedge \partial_{q_1})^{\mu\nu} \bigl\{\delta^4(q_1+q_2)\, \mathcal{A}(P|P+Q) \bigr\}\nonumber\\
    &-i\sum_{X}\int d\mu(w_1,w_2) \ \delta^4(q_1+q_2-w_1-w_2 -\{k_X\})\,\mathcal{A}^*\bigl(P+W,\{k_X\} | P+Q\bigr) \nonumber \\ 
    & \times w_1^{\alpha_1}\cdots w_1^{\alpha_N}((p_1+w_1)\wedge \partial_{w_1})^{\mu\nu} \bigl \{ \delta^4(w_1+w_2+\{k_X\}) \ \mathcal{A}\bigl(P|P+W,\{k_X\}\bigr) \bigr\}\Big] .\label{e5.16}
\end{align}
Once again, the $L$th loop contribution can be written in the same form as for the angular impulse. The linear-amplitude contribution $R'_{(L)}$ is
\begin{align}
	\Aboxed{R'^{\mu\nu\boldsymbol{\alpha}}_{(L)}&=\frac{e^{2L+2}}{4\,\hbar^{L-N}}\int d\mu_{\bar q}\,e^{-ib\cdot\bar q}\sum_{n=0}^L\frac{(-1)^n\,\hbar^n}{2^n\,n!}\,H_{1,n}^{(L)\,\mu\nu\boldsymbol{\alpha}},\hspace{2mm}(L\ge0),}\label{e5.17}
\end{align}
Here the vector index $\boldsymbol{\alpha}=\alpha_1\cdots\alpha_N$. The cut contribution is written as follows,
\begin{align}
    \hspace{10mm}\Aboxed{C'^{\mu\nu\boldsymbol{\alpha}}_{(L)}&=\frac{e^{2L+2}}{4\,\hbar^{L-N}}\int d\mu_{\bar q}\,e^{-ib\cdot\bar q}\sum_{n=0}^L\frac{(-1)^n\,\hbar^n}{2^n\,n!}\,H_{2,n}^{(L)\,\mu\nu\boldsymbol{\alpha}},\hspace{2mm}(L\ge0),}\label{e5.18}
\end{align}
The expressions for $H_{i,n}^{(L)\,\mu\nu\boldsymbol{\alpha}}$ are also given in the Appendix (\ref{A}).

\section{Log soft constraints on radiation kernel}\label{s6}
In this section, we will derive a set of conditions that the ``hard amplitude" has to satisfy so that the radiation kernel produces a tail to the electro-magnetic memory at a generic order in perturbative expansion. One of these two conditions corresponds to the constraint that, in the classical limit, the S matrix satisfies macroscopic causality (retarded propagation). We now outline our proposal to impose macroscopic causality in the context of KMOC observables and then use it as a necessary condition that ensures that the sub-leading soft radiation kernel matches with classical log soft theorem. 

Let $\langle O \rangle$ be an inclusive observable computed using KMOC. It can be written schematically as, 
\begin{align}\label{dec24}
\langle O \rangle &= \sum_{L}\, \int_{\bar q} \, \prod_{i=1}^{L} \hat d^4 l_i \frac{N(p_1,p_2,\bar q,\{l_j\})}{D(p_1,p_2,\bar q,\{l_j\}},
\end{align}
where $1/D$ is the product of propagators. We propose that the requirement of macroscopic causality can be formulated by demanding that $\langle O \rangle$ remains invariant if all the propagators on the RHS of (\ref{dec24}) are replaced by retarded propagators.\footnote{The traditional formulation of macroscopic causality in S-matrix theory by Stapp and Iagolnitzer is a requirement that only cut propagators contribute in the classical limit. While we do believe that the macroscopic causality criteria used in this paper are equivalent to the formulation in \cite{Iagolnitzer:1969sk}, a detailed comparison of the two is beyond the scope of this paper.}

It can be seen that this condition is equivalent to demanding that \emph{one can always choose the contour of intergration in complex $l^{(i)}_{0}$ plane $\forall\, i\, \in\, \{1,\, \dots, L\}$ such that only matter poles contribute.} Hence, our formulation of macroscopic causality in the context of KMOC formalism is equivalent to showing that the contribution of any photon pole in the evaluation of $\langle O \rangle$ vanishes in the classical limit.

In the next section, we show how the above-stated condition for macroscopic causality is one of the constraints that amplitudes have to satisfy, so that the soft radiation kernel matches the tail to the electro-magnetic memory at $O(\log \omega)$.

\subsection{First set of Consistency conditions.} 

\begin{figure}[!h]
\begin{center}
    \includegraphics[scale=0.3]{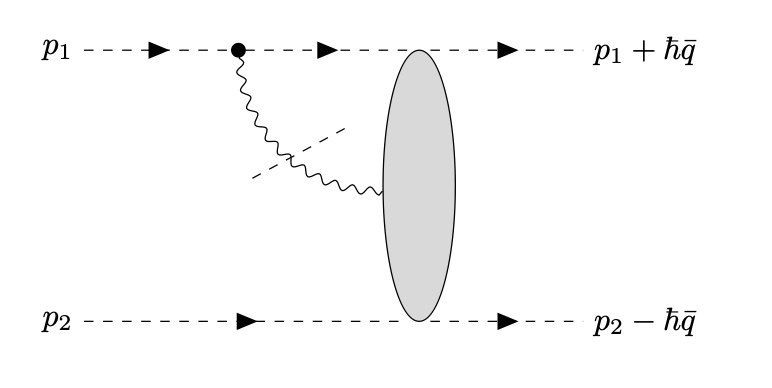}
    \caption{$L$ loop diagram with the outer photon being in the soft region.}
    \label{f3}
    \end{center}
\end{figure}

To calculate the log tail from the sub leading kernel, first we need to extract the $\mathcal{O}(\log \omega)$ contributions from (\ref{e5.7}) and (\ref{e5.8}). At $L$ loop, log divergence arises when the loop momentum of one of the outermost photon lines, say $\bar l_o$, is restricted to the region $\omega < |\bar l_o| \ll |\bar q|$. It turns out such diagrams fall into four classes depending upon how the $l_o$ photon attaches to the $(L-1)$-loop blob. Other diagrams do not produce $\mathcal{O}(\log \omega)$ terms. For example, the diagram (\ref{f3}) does not contribute to $\mathcal{O}(\log \omega)$ as there is only one internal matter propagator and power counting tells us the contribution from these diagrams will be $\mathcal{O}(\omega)$. 

In each of these diagrams, we take the outermost photon leg to be on-shell. If we consider an $L$ loop diagram, then (\ref{f2}) will be the four sets of diagrams that contribute to $\mathcal{O}(\log \omega)$. By taking the outermost photon on-shell, we segregate the photon pole contribution from (\ref{e5.4}). Thus, $\mathcal{R}^{(0)}$ can be written as,
\begin{align}
    \mathcal{R}^{(0)}&=\mathcal{R}^{(0)}_{\text{q}} + \mathcal{R}^{(0)}_{\text{cl}},\label{m6.2}
\end{align}
where $\mathcal{R}^{(0)}_{\text{q}}$ is the quantum term that has photon pole contribution, and $\mathcal{R}_{\text{cl}}^{(0)}$ is the classical contribution coming from the matter poles. The vanishing of the photon pole contribution in the classical limit provides constraints on the perturbative amplitude,
\begin{align}
    \mathcal{R}^{(0)}_{\text{q}} &= 0.\label{e6.3}
\end{align}

$\mathcal{R}^{(0)}_{\text{q}}$ in (\ref{e5.4}) consists of two terms $\Big \langle \Delta \hat J^{\mu \nu}_1 \Big \rangle_{\text{q}}$ and $\Big \langle q_1^{\alpha_1}\cdots q_1^{\alpha_N}\hat J_{1,\text{out}}^{\mu \nu} \Big \rangle_{\text{q}}$ which vanish separately owing to their different tensor structures. 

\subsection{Constraints at L loop}
\begin{figure}[!h]
    \includegraphics[scale=0.3]{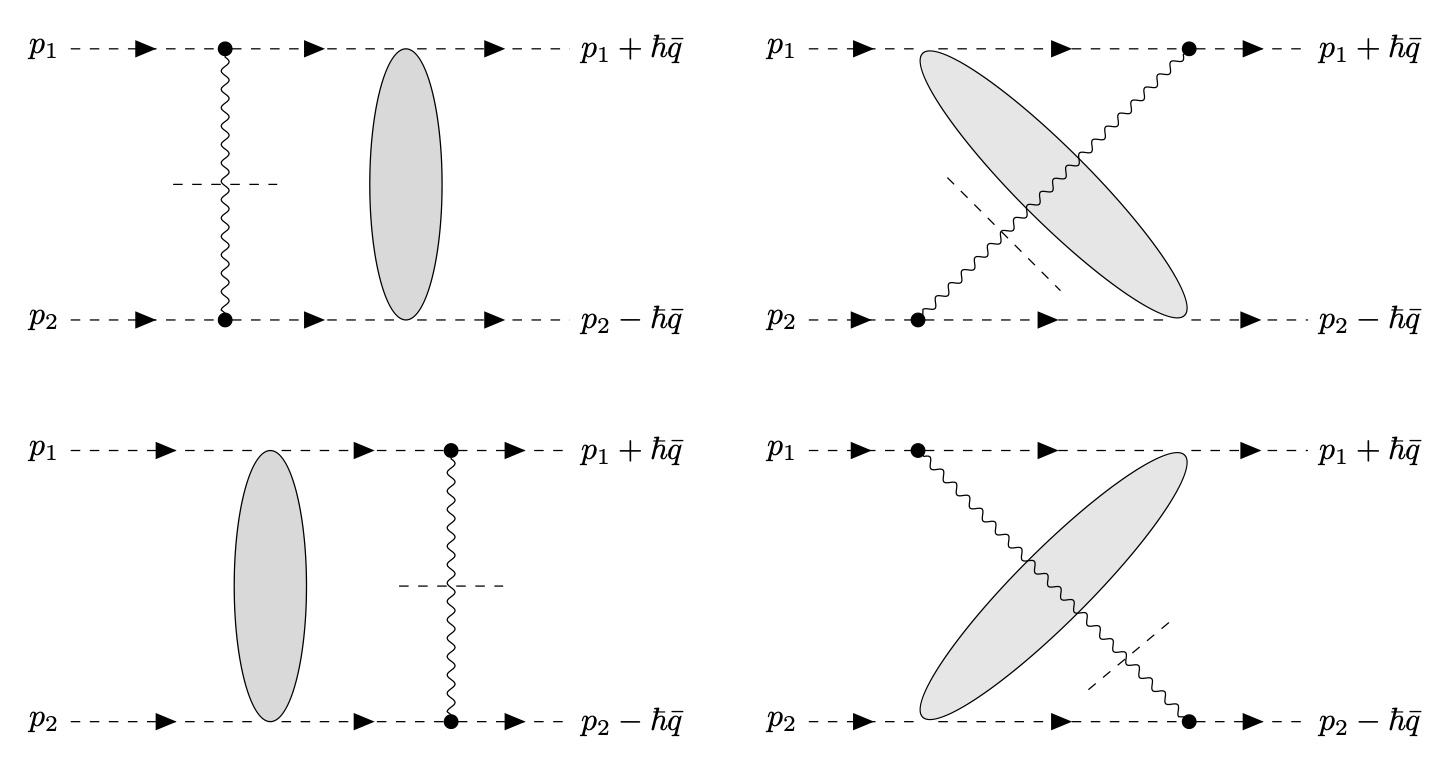}
    \caption{The four classes of $L$ loop diagrams that give log-divergent contributions in the soft region of the photon separated from the $L$ loop.}
    \label{f2}
\end{figure}

We start by calculating the linear amplitude contribution to $\Big \langle \Delta \hat J^{\mu \nu}_1 \Big \rangle_{\text{q}}$ in (\ref{e5.4}),  $R_{(L)\,\text{q}}$, from (\ref{e5.7}). We denote by $\mathcal{\tilde K}_{(L)}$ the sum of the diagrams over all four classes of diagrams without a cut on the outer photon across the $L-1$ loop blob. $\mathcal{\tilde K}_{(L)\,\text{q}}$ then denotes the diagrams with a cut on the outer photon, the sum of diagrams in Fig. (\ref{f2}). The remaining term $\mathcal{\tilde K}_{(L)}-\mathcal{\tilde K}_{(L)\,\text{q}}$ contributes to the $\mathcal{R}_{\text{cl}}^{(0)}$ in (\ref{m6.2}).
\begin{align}
    \mathcal{\tilde K}_{(L)}&=\mathcal{\tilde K}_{(L)\,\text{q}}+\mathcal{\tilde K}_{(L)\,\text{cl}}.\label{e6.04}
\end{align}

Photon poles can be isolated by simply replacing the photon propagator with the delta function $1/\bar l^2\rightarrow \hat\delta(\bar l^2)$ as shown by the cut in Fig. (\ref{f2}). We write the quantum contribution to $R_{(L)}^{\mu \nu}$, using (\ref{e5.7}), 
\begin{align}
    R_{(L),\,\text{q}}^{\mu \nu} &= \frac{1}{4\hbar^L} \int d \mu_{\bar q} \, e^{-ib\cdot \bar q} \sum_{n=0}^L\frac{(-1)^n\hbar^n }{2^n \ n!} (\partial_{x_1}-\partial_{x_2})^n  \nonumber \\
	&\Big[(\bar q^2)^n \ (p_1\wedge \partial_{p_1})^{\mu \nu} \mathcal{\tilde K}_{(L)\,\text{q}}- \partial_{x_1}\Big((\bar q^2)^n \ (p_1\wedge \bar q)^{\mu \nu} \mathcal{\tilde K}_{(L)\,\text{q}}\Big)\Big].\label{e6.1}
\end{align}

We separate the two tensor structures that appear in the above expression. The term $(p_1\wedge \partial_{p_1})^{\mu \nu} \mathcal{\tilde K}_{(L)\,\text{q}}$ give rise to terms of the form,
\begin{align*}
    (p_1 \wedge \partial_{p_1})^{\mu \nu} \mathcal{\tilde K}_{(L)\,\text{q}} &= (p_1 \wedge \bar q)^{\mu \nu} \mathcal{\tilde K}_{(L)\,\text{q}}^{\,(p_1 \wedge \bar q)}+(p_1 \wedge p_2)^{\mu \nu} \mathcal{\tilde K}_{(L)\,\text{q}}^{\,(p_1 \wedge p_2)}.
\end{align*}

Here, $\mathcal{\tilde K}_{(L)\,\text{q}}^{(\,p_1 \wedge \bar q)}$ is the coefficient of $(p_1 \wedge \bar q)^{\mu \nu}$ term in $(p_1 \wedge \partial_{p_1})^{\mu \nu} \mathcal{\tilde K}_{(L)\,\text{q}}$. The intermediate term of the form $(p_1\wedge \bar l)^{\mu \nu}\,\mathcal{\tilde K}_{(L)\,\text{q}}^{\,(p_1\wedge \bar l)}$, where $l$ is the loop momentum, can be tensor-decomposed into the above two terms. $R_{(L),\,\text{q}}^{\mu \nu}$ in (\ref{e6.1}) can be written as
\vspace{-2mm}
\begin{align}
    R_{(L),\,\ln}^{\mu \nu} &= \frac{1}{4\hbar^L} \int d \mu_{\bar q} \, e^{-ib\cdot \bar q} \sum_{n=0}^L\frac{(-1)^n\hbar^n }{2^n \ n!} (\partial_{x_1}-\partial_{x_2})^n  \nonumber \\
	&\Big[(p_1 \wedge \bar q)^{\mu \nu} \,(\bar q^2)^n\,\mathcal{\tilde K}_{(L)\,\text{q}}^{\,(p_1 \wedge \bar q)}- (p_1\wedge \bar q)^{\mu \nu}\,\partial_{x_1}\left((\bar q^2)^n \ \mathcal{\tilde K}_{(L)\,\text{q}}\right)\nonumber\\
    &+(p_1 \wedge p_2)^{\mu \nu} \,(\bar q^2)^n\,\mathcal{\tilde K}_{(L)\,\text{q}}^{\,(p_1 \wedge p_2)}-\frac{p_1\cdot p_2}{D}\,(p_1\wedge p_2)^{\mu \nu}\,(\bar q^2)^n \ \mathcal{\tilde K}_{(L)\,\text{q}}\Big].\label{e6.2}
\end{align}
Next, we turn to calculate the cut contribution. Proceeding in the same spirit, we write the quantum terms in (\ref{e5.8}) as follows
\vspace{-2mm}
\begin{align}
	C_{(L),\,\text{q}}^{\mu\nu}&=\frac{i}{4\hbar^L} \int d \mu_{\bar q} \, e^{-ib\cdot \bar q} \sum_{n=0}^L\frac{(-1)^n\hbar^n }{2^n\, n!} (\partial_{x_1}-\partial_{x_2})^n \left[(\bar q^2)^n\int d \mu(\bar w_1, \bar w_2)\left(-\mathcal{T}_{1\,(L)\,\text{q}}^{\mu\nu}+\mathcal{T}_{2\,(L)\,\text{q}}^{\mu\nu}\right)\right.\nonumber\\
	&- \, 2\,(\bar q^2)^n \int \hat d^4(\bar w_1, \bar w_2)\hat\delta'(2p_1\cdot \bar w_1+\hbar \bar w_1^2)\hat\delta(2p_2 \cdot \bar w_2+\hbar \bar w_2^2)\, (p_1\wedge \bar w_1)^{\mu\nu}\,\mathcal{T}_{3\,(L)\,\text{q}}\nonumber\\
    &\left.2\,\partial_{x_1}\int\! d \mu(\bar w_1,\bar w_2)\, (\bar q^2)^n \,(\bar w_1 \wedge (p_1+\hbar \bar q_1))^{\mu\nu}\,\mathcal{T}_{3\,(L)\,\text{q}}\right].\label{e6.4}
\end{align}
$\mathcal{T}_i$ are the quadratic amplitude terms after integrating out the intermediate photons. The explicit expressions for $\mathcal{T}_i$ are given in the Appendix (\ref{B}). The subscript q on $\mathcal{T}_i$ denotes the photon pole contribution from the diagrams constituting $\mathcal{T}_i$. As with the case of linear amplitude, $C_{(L)}$ gets the log-divergent contribution from a certain class of diagrams. The sum of log-divergent diagrams, $\mathcal{\tilde K}_{(L)}$, is defined in a similar way to the previous case. The momentum of the outermost photon in $\mathcal{\tilde K}_{(L)}$ is restricted to the region $\omega<|\bar l|<<|\bar q|$. One such diagram with a cut on the outer photon is shown below, Fig. (\ref{f4}). The conjugate diagrams are defined in a similar way.
\begin{figure}[!h]
\begin{center}
    \includegraphics[scale=0.22]{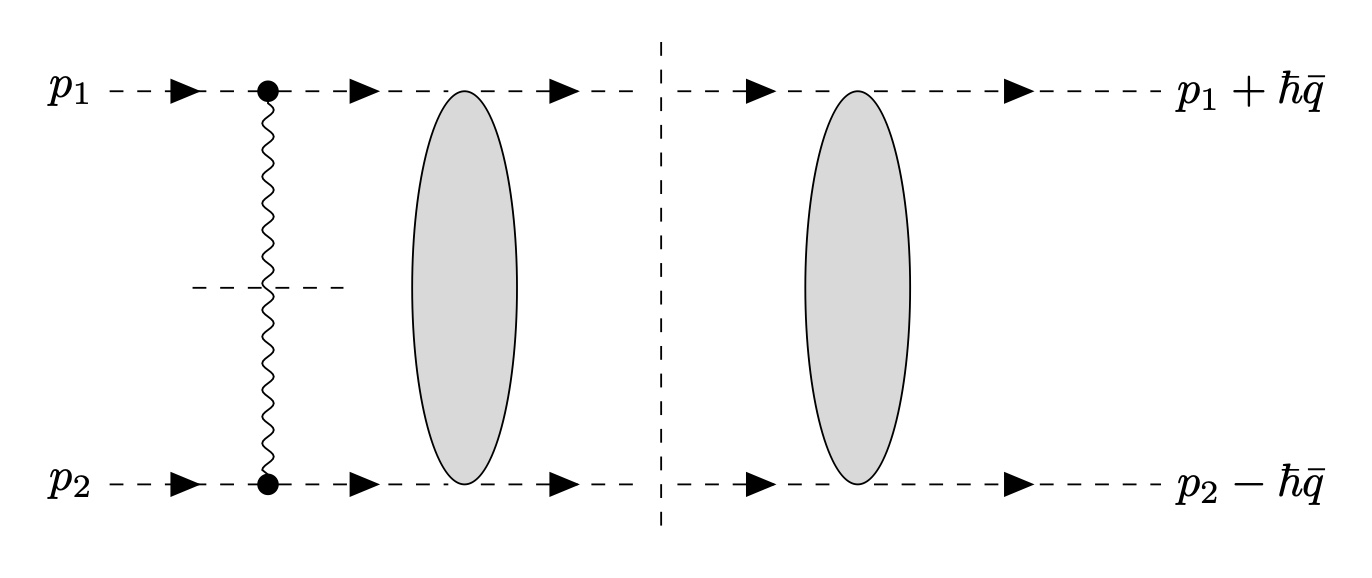}
    \vspace{-6mm}
    \caption{Example of $L$ loop diagram appearing in the cut contribution.}
    \label{f4}
    \end{center}
\end{figure}

Now that we have both linear amplitude and cut contributions, the total contribution from photon poles vanishes in the classical limit, and we have the following constraint,
\begin{align}
	&\Big \langle \Delta \hat J^{\mu \nu}_1 \Big \rangle_{(L),\,\text{q}}=\,\frac{1}{\hbar^L} \int d \mu_{\bar q} \, e^{-ib\cdot \bar q} \sum_{n=0}^L\frac{(-1)^n\hbar^n }{2^n\, n!} (\partial_{x_1}-\partial_{x_2})^n \nonumber\\
	&\Bigg[(p_1\wedge \bar q)^{\mu\nu}\Big( (\bar q^2)^n\, \mathcal{\tilde K}_{(L)\,\text{q}}^{p_1\wedge \bar q}-\partial_{x_1}\left((\bar q^2)^n \ \mathcal{\tilde K}_{(L)\,\text{q}}\right)\Big)\nonumber\\
	&+(p_1\wedge p_2)^{\mu\nu}\,(\bar q^2)^n\left(\mathcal{\tilde K}_{(L)\,\text{q}}^{(p_1\wedge p_2)}-\frac{p_1.p_2}{D}\,\mathcal{\tilde K}_{(L)\,\text{q}}\right)+\,i(\bar q^2)^n\int d\mu(\bar w_1,\bar w_2)\,(-\mathcal{T}_{1\,(L)\,\text{q}}^{\mu\nu}+\mathcal{T}_{2\,(L)\,\text{q}}^{\mu\nu}) \nonumber\\
	&-\,2i(\bar q^2)^n\int \hat d^4(\bar w_1,\bar w_2)\hat\delta'(2p_1\cdot \bar w_1+\hbar \bar w_1^2)\hat\delta(2p_2\cdot \bar w_2+\hbar \bar w_2^2)\,(p_1\wedge \bar w_1)^{\mu\nu}\,\mathcal{T}_{3,(L)\,\text{q}}\nonumber\\
	&+\,2i\,\partial_{x_1}\int d\mu(\bar w_1, \bar w_2) \,(\bar q^2)^n\,(\bar w_1\wedge(p_1+\hbar \bar q))^{\mu\nu}\mathcal{T}_{3,(L)\,\text{q}}\Bigg]=0.\label{e6.8}
\end{align}

The above constraint can further be reduced to three separate constraints. $(R+C)_{(L),\,\ln}^{\mu\nu}$ is antisymmetric, build out of vectors $p_1^{\mu},p_2^{\mu}$ and $b^{\mu}$. Tensor structure of the above expression has the following form
\begin{eqnarray}
	\Big \langle \Delta \hat J^{\mu \nu}_1 \Big \rangle_{(L),\,\text{q}} &=& \mathcal{Z}_1 \, (p_1\wedge p_2)^{\mu\nu}+\mathcal{Z}_2 \, (p_1\wedge b)^{\mu\nu}+\mathcal{Z}_3 \, (p_2\wedge b)^{\mu\nu}.\label{e6.9}
\end{eqnarray}

Here,
\begin{align*}
    \mathcal{Z}_1 &= -\frac{1}{2D}\,\Big \langle \Delta \hat J^{\mu \nu}_1 \Big \rangle_{(L),\,\text{q}} (p_1\wedge p_2)_{\mu\nu}, \quad
	\mathcal{Z}_2= \frac{1}{b^2\,D}\,\Big \langle \Delta \hat J^{\mu \nu}_1 \Big \rangle_{(L),\,\text{q}} (((p_1 \cdot p_2) p_2-m_2^2p_1)\wedge b)_{\mu\nu},\\
	\mathcal{Z}_3&= \frac{1}{b^2\,D}\,\Big \langle \Delta \hat J^{\mu \nu}_1 \Big \rangle_{(L),\,\text{q}} (((p_1 \cdot p_2) p_1-m_1^2p_2)\wedge b)_{\mu\nu}.
\end{align*}

Using the above decomposition, the constraint (\ref{e6.3}) gives
\begin{align}
    \mathcal{Z}_i(p_1,p_2,b;L)&=0.\label{e6.09}
\end{align}

The second constraint is obtained from the vanishing of the photon poles contributions from the second quantity in (\ref{e5.4}). Once again, for the log-divergent contribution in (\ref{e5.16}), we use the diagrams (\ref{f2}) and (\ref{f4}) respectively in (\ref{e5.17}) and (\ref{e5.18}).
\begin{align}
    (R'+C')_{(L),\,\text{q}}^{\mu\nu\boldsymbol{\alpha}}=&-i \frac{e^{2L+2}}{4\,\hbar^{L-N}}\int d\mu_{\bar q}\,e^{-ib\cdot\bar q}\sum_{n=0}^L\frac{(-1)^n\,\hbar^n}{2^n\,n!}\,(\partial_{x_1}-\partial_{x_2})^n\nonumber\\
    &\Bigg[\int d \mu(\bar w_1, \bar w_2)\,\bar w_1^{\,\alpha_1}\cdots\bar w_1^{\,\alpha_N}\,\mathcal{T}^{\mu\nu}_{4\,(L)\,\text{q}}\nonumber\\
    &+\partial_{x_1}\int d \mu(\bar w_1, \bar w_2)\,((p_1+\hbar \bar w_1)\wedge(p_1+\hbar \bar q))^{\mu\nu}\,\bar w_1^{\,\alpha_1}\cdots\bar w_1^{\,\alpha_N}\,\mathcal{T}_{5\,(L)\,\text{q}}\bigg].\label{e6.10}
\end{align}

Again, the expressions for $\mathcal{T}^{\mu\nu}_{4\,(L)}$ and $\mathcal{T}^{\mu\nu}_{5\,(L)}$ are given in the Appendix (\ref{B}). As we see, this does not get any contribution from the linear-amplitude term. We have dropped the terms with tensors $(p_1+\hbar\bar q)^{[\mu}\eta^{\nu]\,(\alpha_1}\bar q^{\alpha_1}\cdots \bar q^{\alpha_N)}$ and $(p_1+\hbar\bar w_1)^{[\mu}\eta^{\nu]\,(\alpha_1}\bar w_1^{\alpha_1}\cdots \bar w_1^{\alpha_N)}$ as these terms vanish on contracting with $\varepsilon_{\mu}k_{\nu}k_{\alpha_1}$ in (\ref{e5.4}). Moreover, the terms of the form $\frac{i}{\hbar}\bigl((p_1+\hbar\bar q)\wedge b\bigr)^{\mu \nu}\,\bar q^{\,\alpha_1}\cdots\bar q^{\,\alpha_N}$ and $\frac{1}{\hbar}\bigl((p_1+\hbar\bar w_1)\wedge b\bigr)^{\mu \nu} \, \bar w_1^{\,\alpha_1}\cdots\bar w_1^{\,\alpha_N}$ combine, when integrated over full region in loop integration, to product of impulses using the result proved in section (\ref{s4}) consequently, they are IR-finite. We obtain the constraint
\begin{align}
    \Big \langle q_1^{\alpha_1}\cdots q_1^{\alpha_N}\hat J_{1,\text{out}}^{\mu \nu} \Big \rangle_{(L),\,\text{q}}&=0.\label{e6.11}
\end{align}
The constraints (\ref{e6.09}) and (\ref{e6.11}) are the vanishing condition in the classical limit for the quantum log terms found in the quantum calculation of the subleading soft factor \cite{Sahoo:2018lxl}.

\subsection{One-loop Calculation}

To calculate $\mathcal{Z}_i$'s, we first calculate the log divergent contribution to (\ref{e6.8}) at $1$-loop. First, we consider the cut contributions. Following from (\ref{eb.1}), (\ref{eb.2}), and (\ref{eb.3}), we see that the amplitudes in these equations are tree-level amplitudes. Cutting one of the photon legs will introduce a delta function, $\hat{\delta}(\bar w^2)$ or $\hat{\delta}((\bar w - \bar q)^2)$.

\begin{center}
    \begin{figure}[!h]
        \includegraphics[scale=0.3]{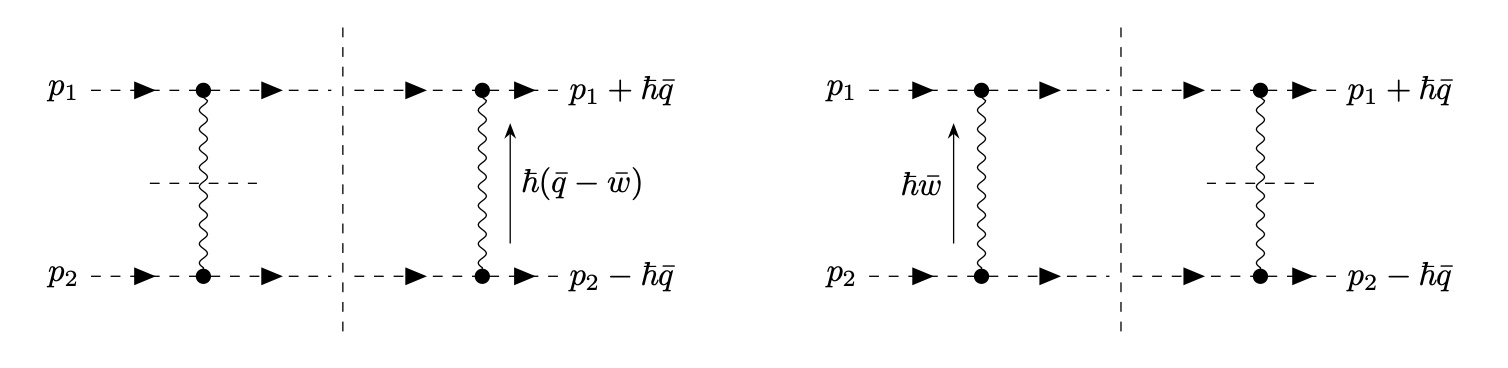}
        \vspace{-13mm}
        \caption{The 1-loop cut diagrams corresponding to the two possibilities of the photon leg being soft.}
        \label{f5}
    \end{figure}
\end{center}

For the first diagram in Fig. (\ref{f5}), $\hat{\delta}(\bar w^2)$ together with the on-shell delta functions $\hat{\delta}(p_i \cdot \bar w)$ makes the $\bar w$ integrals vanish and for the second diagram, $\hat{\delta}((\bar w - \bar q)^2)$ in the relevant region of integration $\omega < |\bar w| \ll |\bar q|$,
becomes $\hat{\delta}(\bar q^2)$. By the same argument, $\bar q$ integrals also vanish. The same argument also shows that the cut contribution to the $\Big \langle q_1^{\alpha_1}\cdots q_1^{\alpha_N}\hat J_{1,\text{out}}^{\mu \nu} \Big \rangle_{(L),\,\text{q}}$in the second constraint (\ref{e6.11}) is zero, ensuring the validity of (\ref{e6.11}).

\begin{center}
    \begin{figure}[!h]
        \includegraphics[scale=0.3]{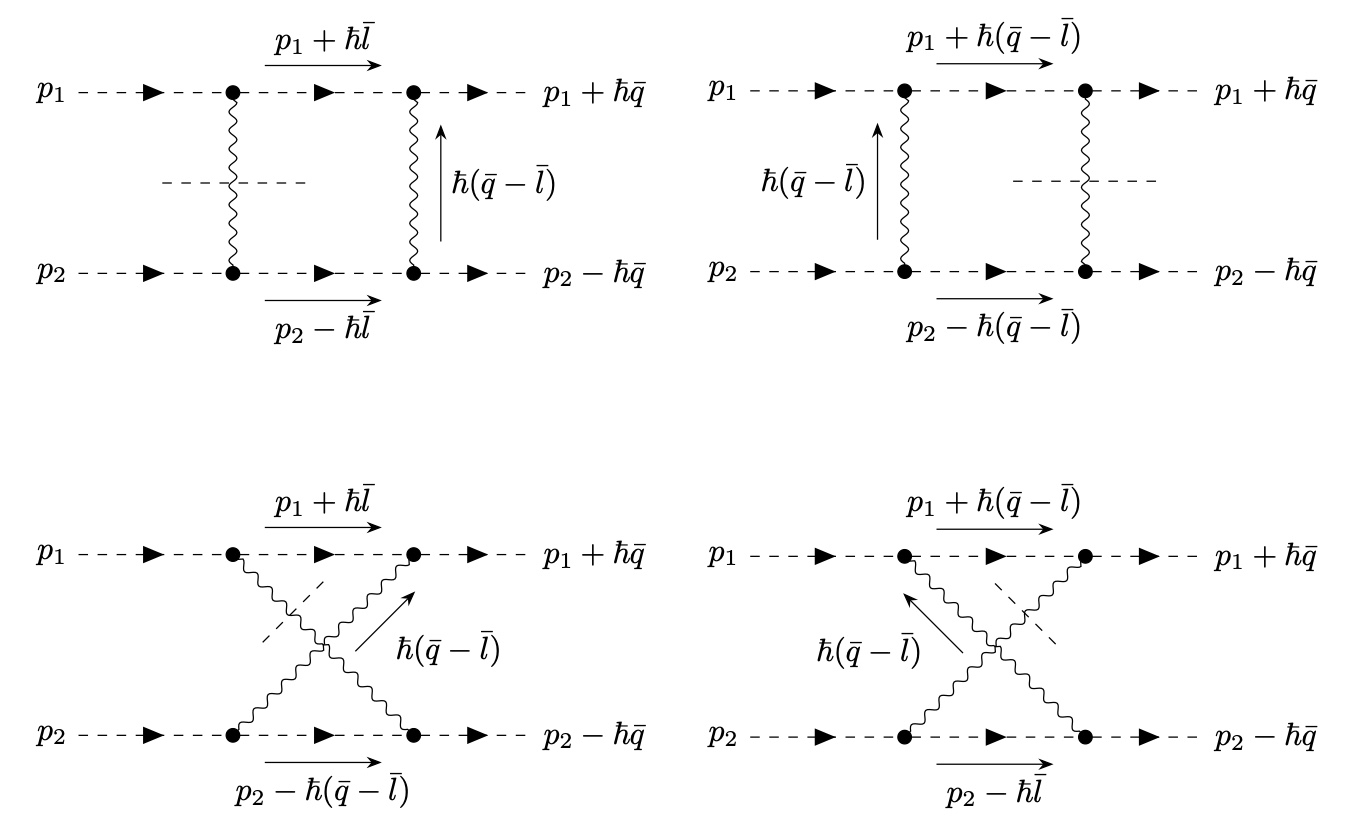}
        \caption{The four 1-loop diagrams corresponding to the four possibilities of the photon leg being soft.}
        \label{f6}
    \end{figure}
\end{center}
Now we turn to calculate the linear contribution in (\ref{e6.8}). The sum over four $1$-loop diagrams, i.e., $\mathcal{\tilde K}_{(1),\,\text{q}}$, has contributions from the following diagrams in Fig. (\ref{f6}). The contribution from these diagrams has been calculated in the Appendix (\ref{C}). The linear contribution has the following three tensor structures, $(p_1\wedge p_2)^{\mu \nu}, \, (p_1\wedge \bar l)^{\mu \nu}, \, (p_1\wedge \bar q)^{\mu \nu}$. At $1$-loop the $(p_1 \wedge \bar l)^{\mu \nu}$ terms give $(p_1 \wedge p_2)^{\mu \nu}$ terms after integration over $\bar l$ while the $(p_1 \wedge \bar q)^{\mu \nu}$ terms give $(p_1 \wedge b)^{\mu \nu}$ terms when integrated over $\bar q$. From the calculation in Appendix (\ref{C}), we have the following results,
\begin{align}
    \mathcal{Z}_2=&\,Q_1^2Q_2^2\frac{(p_1\cdot p_2)^2}{D}\int d \mu_{\bar q} \ e^{-i\bar{q}.b}\ (p_1 \wedge \bar{q})^{\mu \nu}\int\hat d\bar l^4\frac{\hat\delta(\bar l^2)}{(\bar q-\bar l)^2_+} \nonumber\\
    &\times\,\Bigg[\frac{\bar q^2}{(\bar q-\bar l)^2_+}\frac{m_2^2}{(p_1\cdot\bar l)_+}\left(\frac{1}{(p_2\cdot\bar l)_-}-\frac{1}{(p_2\cdot\bar l)_+}\right)\nonumber\\
    &+m_2^2\left(\frac{1}{(p_1\cdot\bar l)_-}-\frac{1}{(p_1\cdot\bar l)_+}\right)\left(\frac{1}{(p_2\cdot\bar l)_-}-\frac{1}{(p_2\cdot\bar l)_+}\right)\nonumber\\
    &+p_1\cdot p_2\left(\frac{1}{(p_1\cdot\bar l)_-}-\frac{1}{(p_1\cdot\bar l)_+}\right)\left(\frac{1}{(p_2\cdot\bar l)_-}-\frac{1}{(p_2\cdot\bar l)_+}\right)\nonumber\\
    &+\frac{\bar q^2}{(\bar q-\bar l)^2_+}\frac{p_1\cdot p_2}{(p_2\cdot\bar l)_-}\left(\frac{1}{(p_1\cdot\bar l)_+}-\frac{1}{(p_1\cdot\bar l)_-}\right)\Bigg].
\end{align}
In the region $\omega<|\bar l|<<|\bar q|$, $\frac{\bar q^2}{(\bar q-\bar l)^2_+}\approx 1+\mathcal{O}(|\bar l|)$. 
\begin{align}
    \mathcal{Z}_2=&\,Q_1^2Q_2^2\frac{(p_1\cdot p_2)^2}{D}\int d \mu_{\bar q} \ e^{-i\bar{q}.b}\ (p_1 \wedge \bar{q})^{\mu \nu}\int\hat d\bar l^4\frac{\hat\delta(\bar l^2)}{(\bar q-\bar l)^2_+} \nonumber\\
    &\times\,\Bigg[\frac{m_2^2}{(p_1\cdot\bar l)_-}\left(\frac{1}{(p_2\cdot\bar l)_-}-\frac{1}{(p_2\cdot\bar l)_+}\right)\nonumber\\
    &-\frac{p_1\cdot p_2}{(p_2\cdot\bar l)_+}\left(\frac{1}{(p_1\cdot\bar l)_-}-\frac{1}{(p_1\cdot\bar l)_+}\right)\Bigg].
\end{align}
$\mathcal{Z}_2$ is zero as there are no log divergent terms present of the form $(p_1\wedge p_2)^{\mu \nu}$ (\ref{C}). $\mathcal{Z}_3$ is also trivially zero as there are no terms that give $(p_2\wedge b)^{\mu \nu}$. The above expression contains integrals of the following type,
\begin{align*}
    \int \hat{d}^4\bar{l} \ \frac{\hat{\delta}(p_1.\bar{l})}{(p_2.\bar{l})_{\pm}}\ \hat{\delta}(\bar{l}^2). 
\end{align*}
This integral vanishes because of the presence of $\hat{\delta}(\bar l^2)$ and on-shell delta functions $\hat{\delta}(p_i \cdot \bar l)$.
%%%%%%%%%%%%%%%%%%%%%%%%%%%%%%%%%%%%%%%%%

\section{Second set of Consistency Conditions}\label{s7}

Having derived the constraints that probe macroscopic causality of the S-matrix thanks to the application of vanishing quantum log soft factor in the KMOC radiation kernel, we now analyze the constraints that \emph{universal classical log soft factor} imposes on a class of conservative observables in the KMOC formalism. These constraints were anticipated in \cite{Bautista:2021llr} and were proved for tree-level scattering in \cite{Manu:2020zxl}.

The subleading contribution to the radiation kernel in (\ref{e5.4}) at $L$ loop can be obtained in a similar manner using similar diagrams as in (\ref{f5}) and (\ref{f6}) without the on-shell cut on the outermost photon line. As with the quantum case, the classical contribution to (\ref{e5.4}) produces two separate constraints.
\begin{eqnarray}
    \langle\Delta\hat J_i^{\mu\nu}\rangle_{\text{cl}} &=& \Delta J_{i,\,{\log}}^{\mu\nu},\\
    \langle q_i^{\alpha_1}\cdots q_i^{\alpha_N}\hat J_{i,\text{out}}^{\mu\nu}\rangle_{\text{cl}} &=& \Delta p_i^{\alpha_1}\cdots \Delta p_i^{\alpha_N} J_{i,\text{out},\,\log}^{\mu\nu},
\end{eqnarray}
with the condition $\mathcal{Z}_i=0$ and $\Big \langle q_1^{\alpha_1}\cdots q_1^{\alpha_N}\hat J_{1,\text{out}}^{\mu \nu} \Big \rangle_{(L),\,\text{q}}=0$. On the right side of the equations are the classical quantities.

As was mentioned previously, the classical contribution comes from the matter poles in the $L$ loop diagram, i.e., from the $\mathcal{\tilde K}_{(L)\,\text{cl}}$ (\ref{e6.04}). Thus, we use $\mathcal{\tilde K}_{(L)\,\text{cl}}$ in (\ref{e5.7}) and (\ref{e5.8}) to isolate the classical contribution to the angular impulse.
\begin{align}
	\langle\Delta\hat J_1^{\mu\nu}\rangle_{(L),\,\text{cl}}=&\,\frac{1}{\hbar^L} \int d \mu_{\bar q} \, e^{-ib\cdot \bar q} \sum_{n=0}^L\frac{(-1)^n\hbar^n }{2^n\, n!} (\partial_{x_1}-\partial_{x_2})^n \nonumber\\
	&\Bigg[(p_1\wedge \bar q)^{\mu\nu}\Big( (\bar q^2)^n\, \mathcal{\tilde K}_{(L)\,\text{cl}}^{p_1\wedge \bar q}-\partial_{x_1}\left((\bar q^2)^n \ \mathcal{\tilde K}_{(L)\,\text{cl}}\right)\Big)\nonumber\\
	&+(p_1\wedge p_2)^{\mu\nu}\,(\bar q^2)^n\left(\mathcal{\tilde K}_{(L)\,\text{cl}}^{(p_1\wedge p_2)}-\frac{p_1.p_2}{D}\,\mathcal{\tilde K}_{(L)\,\text{cl}}\right)\nonumber\\
    &+\,i(\bar q^2)^n\int d\mu(\bar w_1,\bar w_2)\,(-\mathcal{T}_{1\,(L)\,\text{cl}}^{\mu\nu}+\mathcal{T}_{2\,(L)\,\text{cl}}^{\mu\nu}) \nonumber\\
	&-\,2i(\bar q^2)^n\int \hat d^4(\bar w_1,\bar w_2)\hat\delta'(2p_1\cdot \bar w_1+\hbar \bar w_1^2)\hat\delta(2p_2\cdot \bar w_2+\hbar \bar w_2^2)\,(p_1\wedge \bar w_1)^{\mu\nu}\,\mathcal{T}_{3,(L)\,\text{cl}}\nonumber\\
	&+\,2i\,\partial_{x_1}\int d\mu(\bar w_1, \bar w_2) \,(\bar q^2)^n\,(\bar w_1\wedge(p_1+\hbar \bar q))^{\mu\nu}\mathcal{T}_{3,(L)\,\text{cl}}\Bigg].\label{e7.3}
\end{align}

Given $\mathcal{Z}_i(p_1,p_2,b;L)=0$ (\ref{e6.09}), matching the above expression with the classical result \cite{Sahoo:2018lxl} gives the following constraint
\begin{align}
    \langle\Delta\hat J_1^{\mu\nu}\rangle_{(L),\,\text{cl}}&=-\frac{iQ_1Q_2\,m_1^2m_2^2}{4\pi}\,\log\omega\,\Bigg[\frac{(p_1\wedge p_2)^{\mu\nu}}{((p_1\cdot p_2)^2-m_1^2m_2^2)^{3/2}}+\frac{(p'_1\wedge p'_2)^{\mu\nu}}{((p'_1\cdot p'_2)^2-m_1^2m_2^2)^{3/2}}\Bigg]. \label{clcons1}
\end{align}

Another constraint is obtained from the log-divergent classical contribution to\\$\Big\langle q_i^{\alpha_1}\cdots q_i^{\alpha_N}\hat J_{i,\text{out}}^{\mu\nu}\Big\rangle$ in (\ref{e5.4}).

\begin{align}
    \Big\langle q_i^{\alpha_1}\cdots q_i^{\alpha_N}\hat J_{i,\text{out}}^{\mu\nu}\Big\rangle_{(L),\,\text{cl}}=&- \frac{ie^{2L+2}}{4\,\hbar^{L-N}}\int d\mu_{\bar q}\,e^{-ib\cdot\bar q}\sum_{n=0}^L\frac{(-1)^n\,\hbar^n}{2^n\,n!}\,(\partial_{x_1}-\partial_{x_2})^n\nonumber\\
    &\Bigg[\int d \mu(\bar w_1, \bar w_2)\,\bar w_1^{\,\alpha_1}\cdots\bar w_1^{\,\alpha_N}\,\mathcal{T}^{\mu\nu}_{4\,(L)\,\text{cl}}\nonumber\\
    &+\partial_{x_1}\int d \mu(\bar w_1, \bar w_2)\,((p_1+\hbar \bar w_1)\wedge(p_1+\hbar \bar q))^{\mu\nu}\,\bar w_1^{\,\alpha_1}\cdots\bar w_1^{\,\alpha_N}\,\mathcal{T}_{5\,(L)\,\text{cl}}\bigg].\label{e7.5}
\end{align}

The subscript cl on $\mathcal{T}^{\mu\nu}$ denotes the matter pole contribution. We now use the corresponding classical expression to write the constraint
\begin{multline}
     \Big\langle q_i^{\alpha_1}\cdots q_i^{\alpha_N}\hat J_{i,\text{out}}^{\mu\nu}\Big\rangle_{(L),\,\text{cl}} =-\frac{iQ_1Q_2\,m_1^2m_2^2}{4\pi}\,\log\omega\,\frac{(p'_1\wedge p'_2)^{\mu\nu}}{((p'_1\cdot p'_2)^2-m_1^2m_2^2)^{3/2}}\Delta p_1^{\alpha_1}\cdots\Delta p_1^{\alpha_N},\\
    (N \geq 1).\label{e7.6}
\end{multline}
with the condition $\Big\langle q_i^{\alpha_1}\cdots q_i^{\alpha_N}\hat J_{i,\text{out}}^{\mu\nu}\Big\rangle_{(L),\,\text{q}}=0$.

\subsection{One-loop Calculation}

\subsubsection{Contribution from Angular Impulse}

The linear-in-amplitude calculation for the angular impulse follows parallel to the quantum case. We take the outermost photon without an on-shell cut, as the photon pole contribution has already been shown to vanish. We rewrite the expressions obtained in Appendix (\ref{C}), considering only the log-divergent terms in our region of interest.
\begin{align}
    \langle\Delta\hat J_1^{\mu\nu}\rangle_{\text{cl}}=&\,iQ_1^2Q_2^2\frac{(p_1\cdot p_2)^2}{D}\int d \mu_{\bar q} \ e^{-i\bar{q}.b}\ \frac{(p_1 \wedge \bar{q})^{\mu \nu}}{\bar q^2}\int\frac{\hat d\bar l^4}{\bar l^2} \nonumber\\
    &\times\,\Bigg[\frac{m_2^2 \ \hat\delta(p_2\cdot\bar l)}{(p_1\cdot\bar l)_-}-\frac{(p_1\cdot p_2)\,\hat\delta(p_1\cdot\bar l)}{(p_2\cdot\bar l)_+}\Bigg].
\end{align}

It is easy to see that the above integrals vanish. Hence, we see that for the case of one loop, the $\langle\Delta\hat J_1^{\mu\nu}\rangle_{\text{cl}}$ does not get any contribution from the linear amplitude.
The classical log soft factor arises from the $\mathcal{T}_{i\,(1)\,\text{cl}}$ in (\ref{e7.3}) when there is no cut on the photon leg carrying momentum $\bar{w}$.
The cut contribution to (\ref{e7.3}) at $1$-loop can be written as follows,
\begin{align}
    C^{\mu \nu} &= i\hbar \int d\mu(q_1,q_2,w_1,w_2)\ e^{-ib.q_1/\hbar} \delta^4(w_1+w_2-q_1-q_2) \nonumber \\ 
    & \times \mathcal{A}^*(P+Q|P+W) \biggl\{(p_1 \wedge \partial_{p_1})^{\mu \nu} + (w_1 \wedge \partial_{w_1})^{\mu \nu} \biggr\} \bigl(\delta^4(w_1+w_2) \mathcal{A}(P | P+W)\bigr)\label{e7.1}
\end{align}

We substitute the tree-level amplitude $\mathcal{A}_0$ in the above expression. After some algebra, we can write this expression in terms of $C_1^{\mu \nu}$ and $C_2^{\mu \nu}$. $C_1^{\mu \nu}$ consists of the terms where derivative operators act on the tree-level amplitude,
\begin{align}
    C_1^{\mu \nu} &= i\hbar^5 \int \ \prod_{i=1}^{2} \hat{d}^4\bar{q}_i \ \hat{d}^4\bar{w}_i \ \hat{\delta}\bigl(2p_i.\bar{q}_i+\hbar
    \bar{q}_i^{2}\bigr)\hat{\delta}\bigl(2p_i.\bar{w}_i+\hbar
    \bar{w}_i^{2}\bigr) \ e^{-i\bar{q}_1.b} \ \delta^4(\bar{w}_1 + \bar{w}_2 - \bar{q}_1 - \bar{q}_2) \nonumber \\ 
    & \times \delta^4(\bar{w}_1 + \bar{w}_2) \ \mathcal{A}_0^*(P+\bar{Q}|P+\bar{W}) (p_1 \wedge \partial_{p_1})^{\mu \nu}  \bigl( \mathcal{A}_0(P | P+\bar{W})\bigr) \nonumber \\
    &- i\hbar^5 \int \ \prod_{i=1}^{2} \hat{d}^4\bar{q}_i \ \hat{d}^4\bar{w}_i \ \hat{\delta}\bigl(2p_i.\bar{q}_i+\hbar
    \bar{q}_i^{2}\bigr) \hat{\delta}\bigl(2p_i.\bar{w}_i+\hbar
    \bar{w}_i^{2}\bigr) \ e^{-i\bar{q}_1.b} \ \delta^4(\bar{w}_1 + \bar{w}_2 - \bar{q}_1 - \bar{q}_2) \nonumber \\
    & \times \delta^4(\bar{w}_1 + \bar{w}_2) \mathcal{A}_0(P | P+\bar{W})(\bar{w}_1 \wedge \partial_{\bar{w}_1})^{\mu \nu}  \bigl( \mathcal{A}_0^*(P+\bar{Q}|P+\bar{W})\bigr) \\ \nonumber \\
    &= C_{11}^{\mu \nu} + C_{12}^{\mu \nu}\label{e7.7}
\end{align}
$C_2^{\mu \nu}$ contains the terms where derivative operators act on the on-shell delta functions,
    \begin{align}
    C_2^{\mu \nu} &= 2i\hbar^5 \int \ \prod_{i=1}^{2} \hat{d}^4\bar{q}_i \ \hat{d}^4\bar{w}_i \biggl(\hat{\delta}\bigl(2p_1.\bar{q}_1+\hbar
    \bar{q}_1^{2}\bigr) \ \hat{\delta}'\bigl(2p_1.\bar{w}_1+\hbar
    \bar{w}_1^{2}\bigr) + \hat{\delta}'\bigl(2p_1.\bar{q}_1+\hbar
    \bar{q}_1^{2}\bigr) \nonumber \\
    & \times \hat{\delta}\bigl(2p_1.\bar{w}_1+\hbar
    \bar{w}_1^{2}\bigr)\biggr) \hat{\delta}\bigl(2p_2.\bar{q}_2+\hbar
    \bar{q}_2^{2}\bigr) \hat{\delta}\bigl(2p_2.\bar{w}_2+\hbar\bar{w}_2^{2}\bigr) \ (p_1 \wedge \bar{w}_1)^{\mu \nu} e^{-i\bar{q}_1.b} \nonumber \\
    & \times \delta^4(\bar{w}_1 + \bar{w}_2 - \bar{q}_1 - \bar{q}_2) \ \delta^4(\bar{w}_1 + \bar{w}_2) \mathcal{A}_0(P | P+  \bar{W}) \ \mathcal{A}_0^*(P+\bar{Q}|P +  \bar{W}) \nonumber \\
    &+ 2i\hbar^6 \int \ \prod_{i=1}^{2} \hat{d}^4\bar{q}_i \ \hat{d}^4\bar{w}_i \ \hat{\delta}'\bigl(2p_1.\bar{q}_1+\hbar
    \bar{q}_1^{2}\bigr) \hat{\delta}\bigl(2p_1.\bar{w}_1+\hbar
    \bar{w}_1^{2}\bigr) \hat{\delta}\bigl(2p_2.\bar{q}_2+\hbar
    \bar{q}_2^{2}\bigr) \hat{\delta}\bigl(2p_2.\bar{w}_2+\hbar\bar{w}_2^{2}\bigr) \nonumber \\
    & \times (\bar{q}_1 \wedge \bar{w}_1)^{\mu \nu} \ e^{-i\bar{q}_1.b} \ \delta^4(\bar{w}_1 + \bar{w}_2 - \bar{q}_1 - \bar{q}_2) \ \delta^4(\bar{w}_1 + \bar{w}_2) \ \mathcal{A}_0(P | P +  \bar{W}) \nonumber \\ 
    & \times \mathcal{A}_0^*(P+\bar{Q}|P +  \bar{W}) \nonumber \\
    & -i\hbar^5 \int \ \prod_{i=1}^{2} \hat{d}^4\bar{q}_i \ \hat{d}^4\bar{w}_i \ \hat{\delta}\bigl(2p_i.\bar{q}_i+\hbar
    \bar{q}_i^{2}\bigr) \ \hat{\delta}\bigl(2p_i.\bar{w}_i+\hbar
    \bar{w}_i^{2}\bigr) \ e^{-i\bar{q}_1.b} \ \delta^4(\bar{w}_1 + \bar{w}_2 - \bar{q}_1 - \bar{q}_2) \nonumber \\
    & \times \delta^4(\bar{w}_1 + \bar{w}_2) \ \mathcal{A}_0(P | P +  \bar{W}) \ (\bar{w}_1 \wedge \partial_{\bar{w}_1})^{\mu \nu} \ \bigl( \mathcal{A}_0^*(P+\bar{Q}|P +  \bar{W})\bigr) \\
    &= C_{21}^{\mu \nu} + C_{22}^{\mu \nu} + C_{23}^{\mu \nu}
\end{align}

\begin{figure}[!h]
    \begin{center}
    \includegraphics[scale=0.22]{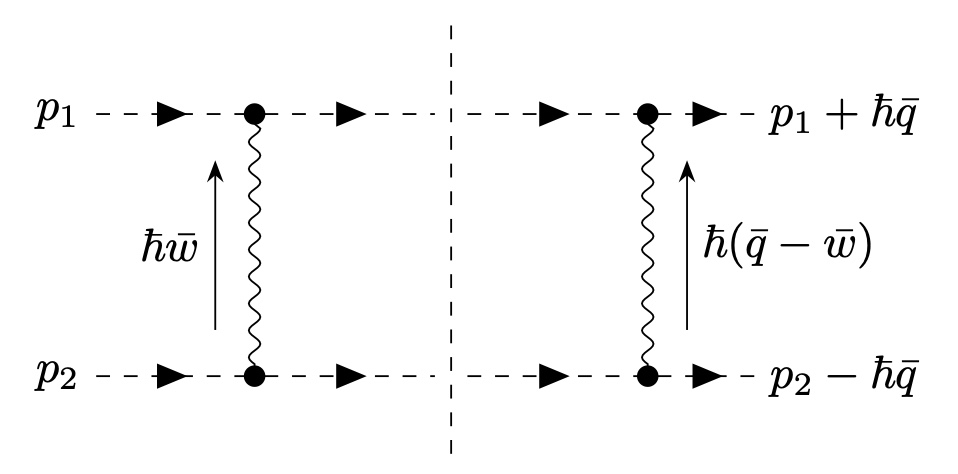}
    \caption{The only possible diagrams for cut-amplitude at 1 loop.}
    \label{f7}
    \end{center}
\end{figure}
The only diagram that contributes in this case is Fig. (\ref{f7}). We will do the transformation $\bar{w} \rightarrow \bar{q}-\bar{w}$ and average over it before evaluating the $\bar{w}$ and $\bar{q}$ integrals. The two tree-level amplitudes in this case are
\begin{align}
    \mathcal{A}_0^*(P+\bar{Q}|P +  \bar{W}) &= \frac{-iQ_1Q_2}{\hbar^3} \biggl(\frac{4p_1.p_2 + 2 \hbar (p_2-p_1).(\bar{w}_1 + \bar{q}_1) + \mathcal{O}(\hbar^2)}{(\bar{q}_2 - \bar{w}_2)^2_+}\biggr),\label{e7.11}\\
    \mathcal{A}_0(P | P +  \bar{W}) &= \frac{iQ_1Q_2}{\hbar^3} \biggl(\frac{4p_1.p_2 + 2 \hbar (p_2-p_1).\bar{w}_1 + \mathcal{O}(\hbar^2)}{(\bar{w}_2)^2_+}\biggr).\label{e7.12}
\end{align}
Using these expressions, we will only write the classical contributions that are $\mathcal{O}(\log\omega)$. The calculation in its entirety has been carried out in the appendix.
\begin{align}
    C_{11}^{\mu \nu} \biggl|_{\log} &= - \frac{iQ_1^2Q_1^2}{4} \ (p_1.p_2) \int d\mu_{\bar q,\bar w}\,e^{-i\bar{q}.b}\frac{(p_1 \wedge \bar{q})^{\mu \nu}}{\bar{w}^2_+ (\bar{q}-\bar{w})^2_+},\label{e7.8}
\end{align}
\begin{align}
    C_{12}^{\mu \nu} \biggl|_{\log} &= \frac{iQ_1^2Q_2^2}{4} (p_1.p_2) \int d\mu_{\bar q,\bar w}\,e^{-i\bar{q}.b}\,\frac{(p_2 \wedge \bar{q})^{\mu \nu}}{\bar{w}^2_+ (\bar{q}-\bar{w})^2_+} .\label{e7.9}
\end{align}
Adding these two contributions, we have,
\begin{align}
    C_1^{\mu \nu} \biggl|_{\log} &= \frac{Q_1Q_2}{8 \pi \sqrt{D}} \bigl((p_2-p_1) \wedge \Delta p\bigr)^{\mu \nu} \ \log\omega.\label{e7.10}
\end{align}
Similarly for $C_2^{\mu \nu}$ we have,
\begin{align}
    C_{21}^{\mu \nu} \biggl|_{\log} &= \frac{3iQ_1^2Q_2^2}{2} (p_1.p_2) \int d\mu_{\bar q,\bar w}\,e^{-i\bar{q}.b}\frac{(p_1 \wedge \bar{w})^{\mu \nu}}{\bar{w}^2_+ (\bar{q}-\bar{w})^2_+}   \nonumber \\
    & - iQ_1^2Q_2^2 \biggl(\frac{m_2^2(p_1.p_2)^2 + (p_1.p_2)^3}{D}\biggr) \int d\mu_{\bar q,\bar w}\,e^{-i\bar{q}.b}\frac{(p_1 \wedge \bar{w})^{\mu \nu}}{\bar{w}^2_+ (\bar{q}-\bar{w})^2_+}  \nonumber \\
    &-iQ_1^2Q_2^2 \biggl(\frac{m_2^2(p_1.p_2)^2 + (p_1.p_2)^3}{D}\biggr)  \int d\mu_{\bar q,\bar w}\, e^{-i\bar{q}.b} \frac{(p_1 \wedge \bar{w})^{\mu \nu}}{((\bar{q}-\bar{w})^2_+)^2}, \label{e7.16}
\end{align}
\begin{align}
    C_{22}^{\mu \nu}\biggl|_{\log} &= iQ_1^2Q_2^2 \frac{m_2^2(p_1.p_2)^2}{D} \int d\mu_{\bar q,\bar w} \,e^{-i\bar{q}.b} \frac{(p_1 \wedge \bar{w})^{\mu \nu}}{\bar{w}^2_+ (\bar{q}-\bar{w})^2_+} \nonumber \\
    & - iQ_1^2Q_2^2 \frac{(p_1.p_2)^3}{D} \int d\mu_{\bar q,\bar w}\,e^{-ib.\bar{q}}  \frac{(p_2 \wedge \bar{w})^{\mu \nu}}{\bar{w}^2_+ (\bar{q}-\bar{w})^2_+},\label{e7.17}
\end{align}
\begin{align}
    C_{23}^{\mu \nu}\biggl|_{\log} &= \frac{iQ_1^2Q_2^2}{2} (p_1.p_2) \int d\mu_{\bar q,\bar w}\,e^{-ib.\bar{q}}\frac{(p_2 \wedge \bar{w})^{\mu \nu}}{\bar{w}^2_+ (\bar{q}-\bar{w})^2_+}.  \label{e7.18}
\end{align}
Finally, adding all these terms we have,
\begin{align}
    C_2^{\mu \nu} \biggl|_{\log} &= \biggl[ \frac{Q_1Q_2}{8 \pi \sqrt{D}} \bigl((3p_1 + p_2) \wedge \Delta p \bigr)^{\mu \nu} - Q_1Q_2 \frac{(p_1 . p_2)^2}{4\pi D^{\frac{3}{2}}} \bigl((p_1+p_2) \wedge \Delta p \bigr)^{\mu \nu} \biggr] \log\omega.\label{e7.14}
\end{align}
And the cut contribution is,
\begin{align}
    C^{\mu \nu} &= -Q_1Q_2 \frac{p_1^2 p_2^2}{4 \pi D^{\frac{3}{2}}} \bigl((p_1+p_2) \wedge \Delta p \bigr)^{\mu \nu} \ \log\omega.\label{e7.15}
\end{align}
This is the angular impulse for $1$ loop as can be seen using the expression for classical angular momentum from \cite{Sahoo:2018lxl},
\begin{align}
    J^{\mu\nu}_a &= -\log\omega \sum_{\stackrel{b\ne a}{\eta_a\eta_b=1}} \frac{Q_aQ_b}{4\pi} \frac{p_a^2 p_b^2\, (p_a\wedge p_b)^{\mu\nu}}{((p_a\cdot p_b)^2 - p_a^2 p_b^2)^{3/2}}.
\end{align}
Here, $a$ and $b$ are the labels for the incoming and outgoing particles. Now for $a=1$ we can calculate the expression for angular impulse at $\mathcal{O}(e^4)$,
\begin{align}
	\Delta J_1^{\mu\nu}|_{\mathcal{O}(e^4)}&= -\log\omega\frac{Q_1Q_2}{4 \pi} \frac{p_1^2 p_2^2\,\bigl((p_1+p_2) \wedge \Delta p \bigr)^{\mu \nu}}{\bigl((p_1 . p_2)^2 - p_1^2 p_2^2\bigr)^{3/2}}.
\end{align}
Hence, (\ref{clcons1}) is verified at one-loop.
\subsubsection{Contribution from \(\Big\langle\frac{(q_1.k)^{N}}{(p_1.k)^{N+1}}\epsilon_{\mu}k_{\nu} J^{\mu \nu}_{1,out}\Big\rangle\)}

The cut contribution for $L$-loop amplitude has the following form,
\begin{align}
    N^{\mu \nu}_{(L)} &= i e^{2L+2} \ \hbar^{N-L-1} \sum_{X=0}^{L-1} \ \sum_{L^{\prime} = 0}^{L-X-1} \biggl[\int \prod_{m=1}^{X}\, d \Phi(k_m) d\mu(\bar q_1,\bar q_2,\bar w_1,\bar w_2) \,e^{-i\bar{q}_1.b}\ \hat{\delta}^4(q_1+q_2)\nonumber \\
    & \times  \ \hat{\delta}^4 \bigr(\bar{w}_1+ \bar{w}_2 + \Sigma_{i=1}^{m} k_i \bigr)\ (\bar{w}_1.k)^N  \ \tilde{\mathcal{A}}_{L^{\prime}}(P | P+\Bar W,\{\bar{k}_X\}) \nonumber \\
    & \times \bigl((p_1 + \hbar \bar{w}_1) \wedge (\partial_{\bar{w}_1}+\partial_{\bar{q}_1}) \bigr)^{\mu \nu} \tilde{\mathcal{A}}^*_{L-L^{\prime}-X-1}(P+\Bar Q | P+\bar{W},\{\bar{k}_X\}) \nonumber \\
    &+ 2\hbar \int \prod_{m=1}^{X} d \Phi(k_m)d\mu(\bar w_1,\bar w_2) \prod_{i=2}^{2} \hat{d}^4 \bar{q}_i \ \hat{\delta}^{\prime}(2p_1.\bar{q}_1 + \hbar \bar{q}_1^2) \ \hat{\delta}(2p_2.\bar{q}_2 + \hbar \bar{q}_2^2) \ e^{-i\bar{q}_1.b}  \nonumber \\
    & \times \ \hat{\delta}^4(q_1+q_2) \hat{\delta}^4 \bigr(\bar{w}_1+ \bar{w}_2 + \Sigma_{i=1}^{m} k_i \bigr)\ (\bar{w}_1.k)^N \ \bigl((p_1+\hbar \bar{w}_1) \wedge (p_1 + \hbar \bar{q}_1) \bigr)^{\mu \nu} \nonumber \\ 
    & \times \tilde{\mathcal{A}}_{L^{\prime}}(P|P+\Bar W,\{\bar{k}_X\}) \, \tilde{\mathcal{A}}^*_{L-L^{\prime}-X-1}(P+\Bar Q|P+\Bar W,\{\bar{k}_X\}) \biggr].\label{e7.19}
\end{align}
In this expression, we have scaled all the couplings and off-shell momenta so that the amplitude starts with $\mathcal{O}(\hbar^0)$ terms. When the operators $\partial_{\bar{w}_1}$ and $\partial_{\bar{q}_1}$ act on $\tilde{\mathcal{A}}^*$, it produces an additional $\hbar$ making the overall scaling as $\hbar^{N-L}$. So for $L$ loop amplitude terms of order $N > L$ will not contribute classically. Now for $1$-loop we write this contribution as $N^{\mu \nu} = N_1^{\mu \nu} + N_2^{\mu \nu}$,
\begin{align}
    N_1^{\mu \nu} &= - i\hbar^5 \int d\mu(\bar q,-\bar q,\bar w, -\bar w)\,e^{-i\bar{q}.b}  \nonumber \\
    &\times  \,\bar{w}. k\,\mathcal{A}_0(P | P+ \bar{W}) ((p_1+\hbar \bar{w}) \wedge \partial_{\bar{w}} )^{\mu \nu} \mathcal{A}_0^*(P+ \bar{W} | P+  \bar{Q}),\label{e7.20}
\end{align}
and
\begin{align}
    N_2^{\mu \nu} &= - i\hbar^5 \int d\mu(\bar q,-\bar q,\bar w,-\bar w)\ e^{-i\bar{q}.b} \nonumber \\
    &\times \bar{w}. k\,  \mathcal{A}_0(P | P+\bar{W}) ((p_1+\hbar \bar{w}) \wedge \partial_{\bar{q}} )^{\mu \nu} \mathcal{A}_0^*(P+ \bar{W} | P+ \bar{Q}) \nonumber \\
    & - 2i\hbar^5 \int d\mu(\bar w,-\bar w) \hat{d}^4\bar{q} \ \hat{d}^4\bar{w} \ \hat{\delta}'\bigl(2p_1.\bar{q}+\hbar
    \bar{q}^{2}\bigr) \ \hat{\delta}\bigl(2p_2.\bar{q}-\hbar
    \bar{q}^{2}\bigr) e^{-i\bar{q}.b}\nonumber \\
    & \times (\bar{w}. k) \bigl((p_1 + \hbar \bar{w}) \wedge (p_1 + \hbar \bar{q})\bigr)^{\mu \nu} \mathcal{A}_0(P | P +  \bar{W}) \mathcal{A}_0^*(P +  \bar{W} | P +  \bar{Q}).\label{e7.21}
\end{align}
The classical contribution from $N_1^{\mu \nu}$,
\begin{align}
    N_1^{\mu \nu} &= - \frac{iQ_1^2Q_2^2}{2}(p_1.p_2)(p_1 \wedge p_2)^{\mu \nu} \int d\mu_{\bar q,\bar w}\,e^{-i\bar{q}.b}\,\frac{\bar{w}.k}{\bar{w}^2_+ (\bar{q}-\bar{w})^2_+}  .\label{e7.22}
\end{align}
Taking the average after the transformation $\bar{w} \rightarrow \bar{q}-\bar{w}$, we have the following,
\begin{align}
    N_1^{\mu \nu} &= - \frac{iQ_1^2Q_2^2}{4}(p_1.p_2)(p_1 \wedge p_2)^{\mu \nu} \int d\mu_{\bar q,\bar w} \,e^{-i\bar{q}.b} \,\frac{\bar{q}.k}{\bar{w}^2_+ (\bar{q}-\bar{w})^2_+} .\label{e7.23}
\end{align}
After evaluating the integrals, we write only $\mathcal{O}(\log\omega)$ terms,
\begin{align}
    N_1^{\mu \nu} &= - \frac{Q_1Q_2}{4 \pi} \frac{\Delta p.k}{2 \sqrt{D}} (p_1 \wedge p_2)^{\mu \nu} \  \log\omega.\label{e7.24}
\end{align}
Similarly, the classical terms from $N_2^{\mu \nu}$,
\begin{align}
    N_2^{\mu \nu} &= - \frac{iQ_1^2Q_2^2}{2}(p_1.p_2)(p_1 \wedge p_2)^{\mu \nu} \int d\mu_{\bar q,\bar w}\,e^{-i\bar{q}.b}\,\frac{\bar{w}.k}{\bar{w}^2_+ (\bar{q}-\bar{w})^2_+} \nonumber \\
    & - 2i Q_1^2Q_2^2 (p_1.p_2)^2 \int d\mu_{\bar w}\,\hat{d}^4\bar{q}\, e^{-i\bar{q}b}\,\hat{\delta}'\bigl(p_1.\bar{q}\bigr)  \hat{\delta}\bigl(p_2.\bar{q}\bigr)\,\frac{\bar{w}.k\, \bigl(p_1 \wedge (\bar{q}-\bar{w})\bigr)^{\mu \nu}}{\bar{w}^2_+ (\bar{q}-\bar{w})^2_+}. \label{e7.25}
\end{align}
Averaging over and doing IBP, we have the following,
\begin{align}
    N_2^{\mu \nu} &= - \frac{iQ_1^2Q_2^2}{4} (p_1.p_2) (p_1 \wedge p_2)^{\mu \nu} \int d\mu_{\bar q,\bar w}\,e^{-i\bar{q}.b}\,\frac{\bar{q}.k}{\bar{w}^2_+ (\bar{q}-\bar{w})^2_+} \nonumber \\
    & + i Q_1^2Q_2^2 \ \frac{(p_1.p_2)^3}{D} (p_1 \wedge p_2)^{\mu \nu} \int d\mu_{\bar q,\bar w}\,e^{-i\bar{q}b} \ \frac{\bar{q}.k}{\bar{w}^2_+ (\bar{q}-\bar{w})^2_+}. \label{e7.26}
\end{align}
Evaluating these integrals we have the $\mathcal{O}(\log\omega)$ terms,
\begin{align}
    N_2^{\mu \nu} &= - \biggl[\frac{Q_1Q_2}{4 \pi} \frac{\Delta p . k}{2 \sqrt{D}} (p_1 \wedge p_2)^{\mu \nu} - \frac{Q_1Q_2}{4 \pi} \frac{(p_1.p_2)^2}{D^{\frac{3}{2}}} (\Delta p .k ) (p_1 \wedge p_2)^{\mu \nu} \biggr] \ \log\omega.\label{e7.27}
\end{align}
Adding these two results, we have,
\begin{align}
    N^{\mu \nu} &= - \frac{Q_1Q_2}{4 \pi} \frac{(\Delta p . k)(p_1^2p_2^2)}{D^{\frac{3}{2}}} (p_1 \wedge p_2)^{\mu \nu} \ \log\omega.\label{e7.28}
\end{align}
Finally, we have the expectation value,
\begin{align}
    \biggl\langle\frac{(q_1.k)}{(p_1.k)^2} \varepsilon_{\mu}k_{\nu} J^{\mu \nu}_{1,out}\biggr\rangle &= - \frac{Q_1Q_2 (p_1^2p_2^2)}{4 \pi D^{\frac{3}{2}}} \frac{\Delta p . k}{(p_1.k)^2} \ \varepsilon_{\mu}k_{\nu} (p_1 \wedge p_2)^{\mu \nu} \ \log\omega.\label{e7.29}
\end{align}
Hence, the second classical constraint \ref{e7.6} is verified at one-loop.

\subsection{Sub-leading soft radiation kernel at $1$-loop}
%\textcolor{red}{Dont write it like this. Write it as constraints on product of impulse observables.}\\
Having calculated the two quantities in the expression (\ref{e5.4}), we can now write the constraints that follow from the universal classical subleading soft factor at one loop. The first of the two constraints is the usual angular impulse at one loop, viz.
\begin{align}
	\langle\Delta \hat J_1^{\mu\nu}\rangle|_{\mathcal{O}(\log \omega,\,e^4)}&= -\frac{Q_1Q_2}{4 \pi} \frac{p_1^2 p_2^2}{\bigl((p_1 . p_2)^2 - p_1^2 p_2^2\bigr)^{\frac{3}{2}}} \bigl((p_1+p_2) \wedge \Delta p \bigr)^{\mu \nu} \ \log\omega.\label{e7.31}
\end{align}
The second constraint is on the product of the first moment and the final angular momentum of the first particle.
\begin{align}
    \langle q^{\alpha}\hat J_1'^{\mu\nu}\rangle|_{\mathcal{O}(\log \omega,\,e^4)} &= - \frac{Q_1Q_2}{4 \pi} \frac{p_1^2 p_2^2}{\bigl((p_1 . p_2)^2 - p_1^2 p_2^2\bigr)^{\frac{3}{2}}}\Delta p^{\alpha}\, (p_1 \wedge p_2)^{\mu \nu} \ \log\omega.\label{e7.32}
\end{align}
\section{Discussion}

Classical soft photon theorems produce a remarkable set of non-perturbative observables for classical electro-magnetic scattering. These observables determine the electro-magnetic memory, and its tails which decay as $\frac{(\ln u)^{m}}{u ^{m+1}}$, where $u$ is the affine parameter on ${\cal I}^{+}$. Universality of the classical soft theorems ensures that these observables only depend on the initial and final momenta, masses, and charges of the scattering particles and are independent of higher-order multipole moments. These theorems have a direct counterpart in S-matrix theory and are known as soft factorization theorems. But even though the electromagnetic memory and the Weinberg soft photon factor are simply ``Fourier transforms" of each other \cite{Strominger:2014pwa}, the tail to the electro-magnetic memory does not equal the Fourier transform of the log soft factor that appears when we expand an amplitude in QED to sub-leading order in the soft expansion.  As argued in \cite{Sahoo:2018lxl, AtulBhatkar:2020hqz}, this distinction is directly tied to the fact that the micro-causality of the quantum S-matrix, which is encapsulated by the nature of the Feynman propagator, is structurally distinct from macroscopic causality of classical scattering which is encapsulated by the fact that the relationship between radiative field and source is governed by the retarded Green's function.  It is reasonable to expect that the classical limit of the S-matrix should satisfy macro-causality, but the precise quantification of this statement is hard to formulate sharply in analytic S-matrix theory, \cite{Iagolnitzer:1969sk}. 

In this paper, we show that one can formulate the macro-causality requirement on the S-matrix of QED in terms of the KMOC formalism. More in detail, we prove that the difference between classical and quantum log soft factors can be used to write a hierarchy of constraints on QED S-matrix so that the radiation kernel computed using the KMOC formalism produce leading classical log soft factor at all orders in the perturbative expansion.  The constraints are summarized in (\ref{e5.5}). These constraints could also be thought of as providing sub-leading soft constraints on KMOC formalism in the spirit of \cite{Bautista:2021llr}, where it was shown that the soft radiative field computed using KMOC formalism equals electro-magnetic memory if and only if a class of observables in the conservative sector satisfy a constraint. In section (\ref{s4}), we prove the leading order soft constraints of \cite{Bautista:2021llr} to all orders in perturbative expansion. 

Our line of inquiry opens up several avenues of investigation. The immediate task is to prove these constraints. While we have verified them at one loop, a proof at all loops remains open. It will also be interesting to know if these constraints can be used as a definition of macro-causality of the S-matrix. Such a proposal would be viable if, for example, the following were true: Assuming that an S-matrix involving photons and charged scalars satisfies (\ref{e5.5}), does it imply that to the sub-subleading order in soft expansion, the radiation kernel equals the tail of tail derived by Sahoo in \cite{Sahoo:2020ryf}? 

Our results suggest that the universality of classical log soft radiation encodes genuinely new, non-perturbative information about the structure of the S-matrix, beyond that captured by standard soft theorems and leading-order memory effects. It would be interesting to investigate whether analogous constraints arise in gravitational scattering, where tail effects and infrared structure are even richer, and to explore whether the quadratic constraint admits an interpretation in terms of symmetry properties of amplitudes.

\section*{Acknowledgments}
We want to express our sincere gratitude to Alok Laddha for suggesting the idea and for his continuous guidance throughout the project. We are also deeply grateful to Sujay Ashok for his valuable insights that greatly helped shape this work. We would further like to thank Samim Akhtar and Akavoor Manu for many helpful discussions and suggestions that contributed to improving this project.
\appendix
\section{Calculation of $F_{i,n}^{(L)},G_{i,n}^{(L)},$ and $H_{i,n}^{(L)}$}\label{A}
We start with the angular impulse in (\ref{e5.6}). We can use IBP and transfer the derivative $\partial_{q_1}$ in the linear amplitude term. One such term, for example, when the derivative acts on the on-shell delta function $\hat\delta(2p_1\cdot q_1+q_1^2)$, is
\begin{align}
   & \frac{2}{\hbar^L}\int \hat d^4(\bar q_1,\bar q_2)\hat\delta'(2p_1\cdot \bar q_1+\hbar\bar q_1^2)\hat\delta(2p_1\cdot \bar q_2+\hbar\bar q_2^2)\hat\delta^4(\bar q_1+\bar q_2)e^{-ib\cdot \bar q_1}p_1\wedge \bar q_1\mathcal{\tilde A}_{(L)}(P|P+\hbar\bar Q).\nonumber
\end{align}
We have scaled the transfer momenta by $\hbar$ and have used the coupling-stripped amplitude $\mathcal{\tilde A}_{(L)}$. Performing $\hbar$ expansion of the on-shell delta functions and, after the change of variable (\ref{e5.9}), transferring the derivatives at each order over to the amplitude, we obtain
\begin{align}
    &-\frac{1}{4\hbar^L}\int d\mu_{\bar{q}} \, e^{-ib\cdot\bar q}\sum_{n=0}^{L}\frac{(-1)^n\hbar^n}{2^n\,n!}\partial_{x_1}(\partial x_1-\partial x_2)^n\left((\bar q^2)^n\,p_1\wedge \bar q_1\,\mathcal{\tilde A}_{(L)}(P|P+\hbar\bar Q)\right),
\end{align}
where $x_i=p_i\cdot\bar q$. We write $F_{i,n}^{(L)}$ in (\ref{e5.7}).
\begin{eqnarray}
	F_{1,n}^{(L)}=&\,&(\partial_{x_1}-\partial_{x_2})^n\Big((\bar q^2)^n\,(p_1\wedge \partial_{p_1}-ib\wedge \bar q)\mathcal{\tilde A}_{(L)}(P|P+\Bar Q)\Big),\label{a.2}\\
	F_{2,n}^{(L)}=&\,& \partial_{x_1}(\partial_{x_1}-\partial_{x_2})^n\Big((\bar q^2)^n\,p_1\wedge\bar q\,\mathcal{\tilde A}_{(L)}(P|P+\Bar Q)\Big).\label{a.3}
\end{eqnarray}
Similarly, we can proceed with the cut contribution in (\ref{e5.6}). We stipulate the result below.
\begin{eqnarray}
	G_{1,n}^{(L)}=&\,&i(\partial_{x_1}-\partial_{ x_2})^n\sum_{X=0}^{L-1}\sum_{L'=0}^{L-X-1}(\bar q^2)^n\!\int\! \prod_{m=1}^Xd\Phi(\bar k_m)\Big[d \mu(\bar w_1, \bar w_2) \hat\delta^4(\bar w_1 + \bar w_2+\Sigma_{i=1}^X \bar k_i) \nonumber\\
	&&\times\,\Big(-p_1\wedge\partial_{p_1} \mathcal{\tilde A}_{(L')}(P| P+\Bar{W},\{\bar k_X\}) \,\, \mathcal{\tilde A}^*_{(L-L'-X-1)}(P+\Bar{Q} |P+\Bar{W},\{\bar k_X\}) \nonumber\,\\
	&&+\,\mathcal{\tilde A}_{(L')}(P|P+\Bar{W},\{\bar k_X\})(\bar w_1 \wedge \partial_{\bar w_1} + \bar q_1 \wedge \partial_{\bar q_1} ) \mathcal{\tilde A}^*_{(L-L'-X-1)}(P+\Bar{Q}|P+\Bar{W}, \{\bar k_X\})\nonumber\\
	&&+i\,b \wedge \bar w_1 \, \mathcal{\tilde A}_{(L')}(P| P+\Bar{W},\{\bar k_X\}) \, \mathcal{\tilde A}^*_{(L-L'-X-1)}(P+\Bar{Q}|P+\Bar{W},\{\bar k_X\})\Big) \nonumber\\
	&&+ \, 2 \int \hat d^4(\bar w_1, \bar w_2)\hat\delta'(2p_1\cdot \bar w_1+\hbar \bar w_1^2)\hat\delta(2p_2 \cdot \bar w_2+\hbar \bar w_2^2) \hat\delta^4(\bar w_1 + \bar w_2 + \Sigma_{i=1}^X \bar k_i) \nonumber\\
	&&\times \, \bar w_1\wedge p_1\,\mathcal{\tilde A}_{(L')}(P|P+\Bar{W},\{\bar k_X\}) \, \, \mathcal{\tilde A}^*_{(L-L'-X-1)}(P+\Bar{Q}|P+\Bar{W},\{\bar k_X\})\Big]_{\bar q_1 = -\bar q_2=\bar q},\label{a.4}\\
	G_{2,n}^{(L)} =&\,& 2i\,\partial_{x_1}(\partial_{x_1}-\partial_{x_2})^n \sum_{X=0}^{L-1} \sum_{L'=0}^{L-X-1} (\bar q^2)^n\!\int\!\prod_{m=1}^Xd\Phi(\bar k_m) d \mu(\bar w_1,\bar w_2) \,\hat\delta^4(\bar w_1 + \bar w_2+\Sigma_{i=1}^X \bar k_i) \nonumber\\
	&&\times\bar w_1 \wedge (p_1+\hbar \bar q_1)\,\mathcal{\tilde A}_{(L')}(P|P+\Bar{W},\{k_X\})\,\mathcal{\tilde A}^*_{(L-L'-X-1)}(P+\Bar{Q}|P+\Bar{W},\{k_X\})\Big|_{\bar q_1= -\bar q_2= \bar q}.\nonumber \label{a.5}
\end{eqnarray}
Calculation of $H_{i,n}^{(L)}$ in (\ref{e5.16}) proceeds in the similar manner.
\begin{align}
    H_{1,n}^{(L)}=& -(\partial_{x_1}-\partial_{x_2})^n\Big((\bar q^2)^n\,(p_1+\hbar \bar q)\wedge\eta^{\cdot\,(\alpha_1}\bar q^{\,\alpha_2}\cdots\bar q^{\,\alpha_N)}\mathcal{\tilde A}_{(L)}(P|P+\Bar Q)\nonumber\\
    &+\frac{i}{\hbar}(p_1+\hbar\bar q)\wedge b\,\bar q^{\,\alpha_1}\cdots\bar q^{\,\alpha_N}\mathcal{\tilde A}_{(L)}(P|P+\Bar Q)\Big),\label{a.6}
\end{align}
and,
\begin{align}
    H_{2,n}^{(L)}=&\,(\partial_{x_1}-\partial_{ x_2})^n\sum_{X=0}^{L-1}\sum_{L'=0}^{L-X-1}(\bar q^2)^n\!\int\! \prod_{m=1}^Xd\Phi(\bar k_m)d \mu(\bar w_1, \bar w_2) \hat\delta^4(\bar w_1 + \bar w_2+\Sigma_{i=1}^X \bar k_i)\nonumber \\
    &\times\,\Big[-i(p_1+\hbar \bar w_1)\wedge\eta^{\cdot\,(\alpha_1}\bar w_1^{\,\alpha_2}\cdots\bar w_1^{\,\alpha_N)}\nonumber\\
    &\hspace{7mm}\times\mathcal{\tilde A}^*_{(L-L'-X-1)}(P+\Bar{Q}|P+\Bar{W},\{\bar k_X\})\mathcal{\tilde A}_{(L')}(P| P+\Bar{W},\{\bar k_X\})\nonumber\\
    &\hspace{7mm}-i\bar w_1^{\,\alpha_1}\cdots\bar w_1^{\,\alpha_N}(p_1+\hbar \bar w_1)\wedge(\partial_{\bar q_1}+\partial_{\bar w_1}) \nonumber\\
    &\hspace{7mm}\mathcal{\tilde A}^*_{(L-L'-X-1)}(P+\Bar{Q}|P+\Bar{W},\{\bar k_X\})\,\mathcal{\tilde A}_{(L')}(P| P+\Bar{W},\{\bar k_X\})\nonumber\\
    &\hspace{7mm}+\frac{1}{\hbar}(p_1+\hbar \bar w_1)\wedge b\,\bar w_1^{\,\alpha_1}\cdots\bar w_1^{\,\alpha_N}\nonumber\\
    &\hspace{7mm}\times\mathcal{\tilde A}^*_{(L-L'-X-1)}(P+\Bar{Q}|P+\Bar{W},\{\bar k_X\})\mathcal{\tilde A}_{(L')}(P| P+\Bar{W},\{\bar k_X\})\Big]_{\bar q_1= -\bar q_2= \bar q}\nonumber\\
    &-2i\partial_{x_1}(\partial_{x_1}-\partial_{ x_2})^n\sum_{X=0}^{L-1}\sum_{L'=0}^{L-X-1}(\bar q^2)^n\!\int\! \prod_{m=1}^Xd\Phi(\bar k_m)d \mu(\bar w_1, \bar w_2) \hat\delta^4(\bar w_1 + \bar w_2+\Sigma_{i=1}^X \bar k_i)\nonumber\\
    &\times\,(p_1+\hbar \bar w_1)\wedge(p_1+\hbar \bar q_1)\bar w_1^{\,\alpha_1}\cdots\bar w_1^{\,\alpha_N}\, \mathcal{\tilde A}^*_{(L-L'-X-1)}(P+\Bar{Q}|P+\Bar{W},\{\bar k_X\})\nonumber\\
    &\times\,\mathcal{\tilde A}_{(L')}(P| P+\Bar{W},\{\bar k_X\})\Big|_{\bar q_1= -\bar q_2= \bar q}.\label{a.7}
\end{align}
\section{Calculating $\mathcal{T}$}\label{B}
In this section, we proceed to write the cut integrals in $\langle\Delta\hat J_i^{\mu\nu}\rangle$ and $\langle q_i^{\alpha_1}\cdots q_i^{\alpha_N}\hat J_{i,\text{out}}^{\mu\nu}\rangle$. The type of diagrams that contribute to the cut (\ref{e6.4}) are defined in the sec (\ref{s6}). We use those diagrams in the expressions (\ref{a.4}) and (\ref{a.5}) and define the following terms
\begin{align}
	\mathcal{T}_{1\,(L)}^{\mu\nu}=&\,\sum_{X=0}^{L-1}\sum_{L'=0}^{L-X-1}\!\int\! \prod_{m=1}^Xd\Phi(\bar k_m)\hat\delta^4(\bar w_1+\bar w_2+\Sigma_{i=1}^X \bar k_i)\,\Big[(p_1\wedge\partial_{p_1})^{\mu\nu}\mathcal{\tilde K}_{(L')}(P|P+\Bar{W},\{\bar k_X\})\,\nonumber\\
	&\times\, \mathcal{\tilde A}^*_{(L-L'-X-1)}(P+\Bar{Q}|P+\Bar{W},\{\bar k_X\}) + (p_1\wedge\partial_{p_1})^{\mu\nu}\mathcal{\tilde A}_{(L')}(P|P+\Bar{W},\{\bar k_X\})\,\nonumber\\
	&\times\, \mathcal{\tilde K}^*_{(L-L'-X-1)}(P+\Bar{Q}|P+\Bar{W},\{\bar k_X\})\Big]_{\bar q_1=-\bar q_2=\bar q},\label{eb.1}
\end{align}

\begin{align}
	\mathcal{T}_{2\,(L)}^{\mu\nu}=&\,\sum_{X=0}^{L-1}\sum_{L'=0}^{L-X-1}\!\int\! \prod_{m=1}^Xd\Phi(\bar k_m)\hat\delta^4(\bar w_1+\bar w_2+\Sigma_{i=1}^X \bar k_i)\,\Big[\mathcal{\tilde K}_{(L')}(P|P+\Bar{W},\{\bar k_X\})\nonumber\\
	&\times\, \big(\bar w_1 \wedge (\partial_{\bar w_1} + \partial_{\bar q_1} )\big)^{\mu\nu}\mathcal{\tilde A}^*_{(L-L'-X-1)}(P+\Bar{Q}|P+\Bar{W},\{\bar k_X\}) +\mathcal{\tilde A}_{(L')}(P|P+\Bar{W},\{\bar k_X\})\nonumber\\
	&\times\, \big( \bar w_1 \wedge (\partial_{\bar w_1} + \partial_{\bar q_1} )\big)^{\mu\nu}\mathcal{\tilde K}^*_{(L-L'-X-1)}(P+\Bar{Q}|P+\Bar{W},\{\bar k_X\})\Big]_{\bar q_1=-\bar q_2=\bar q},\label{eb.2}
\end{align}
and 
\begin{eqnarray}
	\mathcal{T}_{3\,(L)}=&\,&\sum_{X=0}^{L-1}\sum_{L'=0}^{L-X-1}\!\int\! \prod_{m=1}^Xd\Phi(\bar k_m)\,\hat\delta^4(\bar w_1+\bar w_2+\Sigma_{i=1}^X \bar k_i)\,\Big[\mathcal{\tilde K}_{(L')}(P|P+\Bar{W},\{\bar k_X\})\nonumber\\
	&&\times\,\mathcal{\tilde A}^*_{(L-L'-X-1)}(P+\Bar{Q}|P+\Bar{W},\{\bar k_X\})+\mathcal{\tilde A}_{(L')}(P|P+\Bar{W},\{\bar k_X\})\nonumber\\
	&&\times\,\mathcal{\tilde K}^*_{(L-L'-X-1)}(P+\Bar{Q}|P+\Bar{W},\{\bar k_X\})\Big]_{\bar q_1=-\bar q_2=\bar q}.\label{eb.3}
\end{eqnarray}

We also define similar terms for the cut (\ref{e6.10}).
\begin{align}
    \mathcal{T}^{\mu\nu}_{4\,(L)}=&\,\sum_{X=0}^{L-1}\sum_{L'=0}^{L-X-1}(\bar q^2)^n\!\int\! \prod_{m=1}^Xd\Phi(\bar k_m) \hat\delta^4\left(\bar w_1 + \bar w_2+\Sigma_{i=1}^X \bar k_i\right)\,((p_1+\hbar \bar w_1)\wedge(\partial_{\bar q_1}+\partial_{\bar w_1}))^{\mu\nu}\nonumber\\
    &\Big( \mathcal{\tilde K}^*_{(L-L'-X-1)}(P+\Bar{Q}|P+\Bar{W},\{\bar k_X\})\,\mathcal{\tilde A}_{(L')}(P| P+\Bar{W},\{\bar k_X\})\nonumber\\
    &+ \, \mathcal{\tilde A}^*_{(L-L'-X-1)}(P+\Bar{Q}|P+\Bar{W},\{\bar k_X\})\,\mathcal{\tilde K}_{(L')}(P| P+\Bar{W},\{\bar k_X\})\Big)_{\bar q_1= -\bar q_2= \bar q},\label{eb.4}
\end{align}
\begin{align}
    \mathcal{T}_{5\,(L)}=&\,2\sum_{X=0}^{L-1}\sum_{L'=0}^{L-X-1}(\bar q^2)^n\!\int\! \prod_{m=1}^Xd\Phi(\bar k_m)\hat\delta^4(\bar w_1 + \bar w_2+\Sigma_{i=1}^X \bar k_i)\nonumber\\
    &\times\,\Big( \mathcal{\tilde K}^*_{(L-L'-X-1)}(P+\Bar{Q}|P+\Bar{W},\{\bar k_X\})\,\mathcal{\tilde A}_{(L')}(P| P+\Bar{W},\{\bar k_X\})\nonumber\\
    &+\,\mathcal{\tilde A}^*_{(L-L'-X-1)}(P+\Bar{Q}|P+\Bar{W},\{\bar k_X\})\,\mathcal{\tilde K}_{(L')}(P| P+\Bar{W},\{\bar k_X\})\Big)_{\bar q_1= -\bar q_2= \bar q}.\label{eb.5}
\end{align}

\section{$R^{\mu\nu}$ at one loop}\label{C}
In order to extract the classical pieces, we scale the loop momenta and $q$ by $\hbar$ in the diagrams (\ref{f4}). The terms that give finite contributions in the classical limit are $O(\hbar^0)$ and $O(\hbar)$ in the amplitude. We write both terms separately.
\begin{align}
    \mathcal{\tilde B}_{0} + \mathcal{\tilde C}_{0} =&\, 4iQ_1^2Q_2^2 (p_1.p_2)^2 \int \frac{\hat{d}^4 \bar{l}}{\bar{l}^2_+ (\bar{q}-\bar{l})^2_+} \biggl(\frac{1}{(p_1.\bar{l})_+ (p_2.\bar{l} )_-} - \frac{1}{(p_1.\bar{l})_+ (p_2.(\bar{q}-\bar{l}))_+} \nonumber \\ 
    &+ \frac{1}{(p_1.(\bar{q}-\bar{l}))_- (p_2.(\bar{q}-\bar{l}))_+} - \frac{1}{(p_1.(\bar{q}-\bar{l}))_- (p_2.\bar{l} )_-}\biggr).\label{b.1}
\end{align}
\begin{align}
    \mathcal{\tilde B}_1 + \mathcal{\tilde C}_1 =&\, -i\hbar\,Q_1^2Q_2^2 \biggl[ (p_1.p_2) \int \frac{\hat{d}^4 \bar{l}}{\bar{l}^2_+ (\bar{q}-\bar{l})^2_+} \biggl(\frac{6(p_1-p_2).\bar{q}+4(p_2-p_1).\bar{l}}{(p_1.(\bar{q}-\bar{l}))_- (p_2.(\bar{q}-\bar{l}))_+} \nonumber \\
    &+ \frac{2(p_1-p_2).\bar{q}+4(p_1-p_2).\bar{l}}{(p_1.\bar{l})_+ (p_2.\bar{l})_-} + \frac{(2p_2-6p_1).\bar{q} + 4(p_1+p_2).\bar{l}}{(p_1.\bar{l})_+ (p_2.(\bar{q}-\bar{l}))_+} \nonumber \\
    &+ \frac{(6p_2-2p_1).\bar{q}-4(p_1+p_2).\bar{l}}{(p_1.(\bar{q}-\bar{l}))_- (p_2.\bar{l})_-}\biggr) - 2(p_1.p_2)^2 \int \frac{\hat{d}^4\bar{l}}{(\bar{q}-\bar{l})^2_+} \biggl( \frac{1}{(p_1.\bar{l})_+ (p_2.\bar{l})^2_-} \nonumber \\
    &+ \frac{1}{(p_1.\bar{l})^2_+ (p_2.(\bar{q}-\bar{l}))_+} -\frac{1}{(p_1.\bar{l})^2_+ (p_2.\bar{l})_-} - \frac{1}{(p_1.(\bar{q}-\bar{l}))_- (p_2.\bar{l})^2_-} \biggr) \nonumber \\
    &- 2(p_1.p_2)^2 \int \frac{\hat{d}^4 \bar{l}}{\bar{l}^2_+} \biggl(\frac{1}{(p_1.(\bar{q}-\bar{l}))^2_- (p_2.(\bar{q}-\bar{l}))_+} - \frac{1}{(p_1.(\bar{q}-\bar{l}))_- (p_2.(\bar{q}-\bar{l}))^2_+} \nonumber \\
    &+ \frac{1}{(p_1.\bar{l})_+ (p_2.(\bar{q}-\bar{l}))^2_+} - \frac{1}{(p_1.(\bar{q}-\bar{l}))^2_- (p_2.\bar{l})_-}\biggr) \biggl].\label{b.2}
\end{align}
$\mathcal{\tilde B}$ and $\mathcal{\tilde C}$ respectively the box and crossed-box diagrams in Fig. (\ref{f4}). In the intermediate step, we have terms with tensor $(p_1 \wedge \bar l)^{\mu \nu}$. As discussed before, this can be decomposed and absorbed into the two tensor types $(p_1 \wedge \bar p_2)^{\mu \nu}$ and $(p_1 \wedge \bar q)^{\mu \nu}$ to obtain the tensor split written in (\ref{e6.9}). For the purpose of our calculation at 1 loop, we do not perform such a decomposition, as it turns out at 1 loop, considering terms of the form $(p_1 \wedge \bar l)^{\mu \nu}$ separately is sufficient. So we write,
\begin{align}
    R_{(1)}^{\mu \nu} =&\, \frac{i}{4\hbar} \int \hat d^4 \bar q \ \hat\delta(p_1\cdot \bar q) \hat\delta(p_2\cdot \bar q) e^{-ib\cdot \bar q}\Big[(p_1\wedge \partial_{p_1})^{\mu \nu} (\mathcal{\tilde B}_{1} + \mathcal{\tilde C}_{1})- \partial_{x_1}\Big((p_1\wedge \bar q)^{\mu \nu} (\mathcal{\tilde B}_{1} + \mathcal{\tilde C}_{1})\Big)\nonumber\\
    &-\frac{\hbar}{2}(\partial_{x_1}-\partial_{x_2})\Big(\bar q^2 \ (p_1\wedge \partial_{p_1})^{\mu \nu} (\mathcal{\tilde B}_{0} + \mathcal{\tilde C}_{0})- \partial_{x_1}\Big(\bar q^2 \ (p_1\wedge \bar q)^{\mu \nu} (\mathcal{\tilde B}_{0} + \mathcal{\tilde C}_{0})\Big)\Big)\Big].\label{b.3}
\end{align}
The linear contribution can be further simplified as follows,
\begin{align}
    R^{\mu \nu}_{(1)} =&\, \frac{i}{4\hbar} \int \hat{d}^4\bar{q} \ \hat\delta(p_1\cdot \bar q) \hat\delta(p_2\cdot \bar q) \ e^{-i\bar{q}b} \ (p_1 \wedge \partial_{p_1})^{\mu \nu} (\mathcal{\tilde B}_1 + \mathcal{\tilde C}_1) \nonumber \\
    &- \frac{i}{8} \int \hat{d}^4\bar{q} \ \hat\delta(p_1\cdot \bar q) \hat\delta(p_2\cdot \bar q) \ e^{-i\bar{q}b} \ (\partial_{x_1}-\partial_{x_2})(p_1 \wedge \partial_{p_1})^{\mu \nu} (\mathcal{\tilde B}_{0} + \mathcal{\tilde C}_{0}) \nonumber \\
    &- \frac{i}{4\hbar} \int \hat{d}^4\bar{q} \ \hat\delta(p_1\cdot \bar q) \hat\delta(p_2\cdot \bar q) \ e^{-i\bar{q}b} \ \partial_{x_1}\left((p_1 \wedge \bar{q})^{\mu \nu} (\mathcal{\tilde B}_1 + \mathcal{\tilde C}_1)\right) \nonumber \\
    &+ \frac{i}{8} \int \hat{d}^4\bar{q} \ \hat\delta(p_1\cdot \bar q) \hat\delta(p_2\cdot \bar q) \ e^{-i\bar{q}b} \ (\partial_{x_1}-\partial_{x_2})\partial_{x_1}\left((p_1 \wedge \bar{q})^{\mu \nu} (\mathcal{\tilde B}_{0} + \mathcal{\tilde C}_{0})\right) \nonumber \\
    =&\, R_{1,\mathcal{\tilde B}_1 + \mathcal{\tilde C}_1}^{\mu \nu} + R_{1,\mathcal{\tilde B}_{0} + \mathcal{\tilde C}_{0}}^{\mu \nu} + R_{2,\mathcal{\tilde B}_1 + \mathcal{\tilde C}_1}^{\mu \nu} + R_{2,\mathcal{\tilde B}_{0} + \mathcal{\tilde C}_{0}}^{\mu \nu}.\label{b.4}
\end{align}
We consider the three tensors in $R^{\mu\nu}_{(1)}$, $(p_1 \wedge p_2)^{\mu \nu}$, $(p_1 \wedge \bar{q})^{\mu \nu}$, $(p_1 \wedge \bar{l})^{\mu \nu}$, separately.

\begin{align}
    \bigl(R_{1,\mathcal{\tilde B}_1 + \mathcal{\tilde C}_1}^{\mu \nu}\bigr)_{(p_1 \wedge p_2)} =& -iQ_1^2Q_2^2 (p_1 \cdot p_2) (p_1 \wedge p_2)^{\mu \nu} \int d\mu_{\bar{q}} \ e^{-i\bar{q}.b}\int \hat{d}^4\bar{l} \nonumber\\
    &\times \biggl[ \frac{\hat\delta(p_2\cdot\bar l)}{\bar l^2_+ (p_1\cdot\bar l)_-^2}+\frac{\hat\delta(p_1\cdot\bar l)}{\bar l^2_+(p_2\cdot\bar l)_+^2}+\frac{\hat\delta(p_2\cdot\bar l)}{(\bar q-\bar l)^2_+ (p_1\cdot\bar l)_+^2}+\frac{\hat\delta(p_1\cdot\bar l) }{(\bar q-\bar l)^2_+(p_2\cdot\bar l)_-^2}\biggr].\label{b.5}
\end{align}

Next, we have,

\begin{align}
    \bigl(R_{1,\mathcal{\tilde B}_1 + \mathcal{\tilde C}_1}^{\mu \nu}\bigr)_{(p_1 \wedge \bar{q})} =& - \frac{Q_1^2Q_2^2}{2} (p_1.p_2) \int d\mu_{\bar{q}} \ e^{-i\bar{q}.b} (p_1 \wedge \bar{q})^{\mu \nu} \int \frac{\hat{d}^4\bar{l}}{\bar{l}^2_+ (\bar{q}-\bar{l})^2_+} \nonumber \\ 
    & \times\left[-\left(\frac{1}{(p_1\cdot\bar l)_+ (p_2\cdot\bar l)_-}+\frac{1}{(p_1\cdot\bar l)_- (p_2\cdot\bar l)_+}\right)\right.\nonumber\\
    &\left.+\,3\left(\frac{1}{(p_1\cdot\bar l)_+ (p_2\cdot\bar l)_+}+\frac{1}{(p_1\cdot\bar l)_- (p_2\cdot\bar l)_-}\right)\right]\nonumber\\
    &-Q_1^2Q_2^2 (p_1.p_2)^2 \int d\mu_{\bar q} \ e^{-i\bar{q}.b} (p_1 \wedge \bar{q})^{\mu \nu}\int\hat d^4\bar l \nonumber \\
    & \times \left[\frac{i\hat{\delta}(p_2.\bar{l})}{\bar l^2 (p_1\cdot\bar l)_-^3}+\frac{1}{2 \bar l^2 (p_1\cdot\bar l)_-^2 (p_2\cdot\bar l)_+^2}+\frac{1}{2 (\bar q-\bar l)^2 (p_1\cdot\bar l)_-^2 (p_2\cdot\bar l)_-^2}\right].\label{b.6}
\end{align}
Finally, the $(p_1 \wedge \bar{l})$ term, 

\begin{align}
    \bigl(R_{1,\mathcal{\tilde{B}}_1 + \mathcal{\tilde C}_1}^{\mu \nu}\bigr)_{(p_1 \wedge \bar{l})} =& \,iQ_1^2Q_2^2 (p_1.p_2)^2 \int d\mu_{\bar q} \ e^{-i\bar{q}.b} \int \hat{d}^4\bar{l} \, (p_1 \wedge \bar{l})^{\mu \nu} \nonumber \\ 
    &\times \biggl[\biggl(-\frac{\hat{\delta}(p_2.\bar{l})}{\bar l^2_+(p_1.\bar{l})^3_-}- \frac{\hat{\delta}(p_2.\bar{l})}{(\bar q-\bar l)^2_+(p_1.\bar{l})^3_+}  + \frac{\hat{\delta}'(p_1.\bar{l})}{2(\bar q-\bar l)^2_+(p_2.\bar{l})^2_-}  + \frac{\hat{\delta}'(p_1.\bar{l})}{2\bar l^2_+(p_2.\bar{l})^2_+}  \biggr) \biggr].\label{b.7}
\end{align}
Moving onto $\bigl(R_{1,\mathcal{\tilde B}_{0} + \mathcal{\tilde C}_{0}}^{\mu \nu}\bigr)$ we have,

\begin{align}
    \bigl(R_{1,\mathcal{\tilde B}_{0} + \mathcal{\tilde C}_{0}}^{\mu \nu}\bigr)_{(p_1 \wedge p_2)} =&\, Q_1^2Q_2^2 (p_1.p_2)(p_1 \wedge p_2)^{\mu \nu} \int d\mu_{\bar q} \ \bar{q}^2\,e^{-i\bar{q}.b} \int \frac{\hat{d}^4\bar{l}}{\bar{l}^2_+ (\bar{q}-\bar{l})^2_+} \nonumber \\
    & \times \biggl[\frac{(p_1 \cdot p_2) m_1^2 }{2 D }\frac{i\hat\delta(p_1\cdot\bar l)}{(p_1\cdot\bar l)_-^2}+\frac{i\hat\delta(p_2\cdot\bar l)}{(p_1\cdot\bar l)_-^2}+\frac{i\hat\delta(p_1\cdot\bar l)}{(p_2\cdot\bar l)_+^2}+\frac{(p_1 \cdot p_2)^2}{2D}\frac{i\hat\delta(p_2\cdot\bar l)}{(p_1\cdot\bar l)_-^2}\biggr].\label{b.8}
\end{align}
The $(p_1 \wedge \bar{l})^{\mu \nu}$ terms are following,

\begin{align}
    \bigl(R_{1,\mathcal{\tilde B}_{0} + \mathcal{\tilde C}_{0}}^{\mu \nu}\bigr)_{(p_1 \wedge \bar{l})} =&\, Q_1^2Q_2^2 \biggl(\frac{m_2^2(p_1.p_2)^2 + (p_1.p_2)^3}{D}\biggr) \int d\mu_{\bar{q}} \ \bar{q}^2 \ e^{-i\bar{q}.b} \nonumber \\
    & \times \int \frac{\hat{d}^4\bar{l}}{\bar{l}^2_+ ((\bar{q}-\bar{l})^2_+)^2}(p_1 \wedge \bar{l})^{\mu \nu} \hat{\delta}(p_1.\bar{l}) \hat{\delta}(p_2.\bar{l}) + iQ_1^2Q_2^2 (p_1.p_2)^2 \int d \mu_{\bar{q}} \ \bar{q}^2 e^{-i\bar{q}.b} \nonumber \\
    & \times \int \frac{\hat{d}^4\bar{l}}{\bar{l}^2_+ (\bar{q}-\bar{l})^2_+} (p_1 \wedge \bar{l})^{\mu \nu} \biggl[-\frac{\hat{\delta}'(p_1.\bar{l})}{2(p_2.\bar{l})^2_+} + \frac{\hat{\delta}(p_2.\bar{l})}{(p_1.\bar{l})^3_-} \biggr].\label{b.9}
\end{align}

Contributions from terms containing the $(p_1 \wedge \bar{q})^{\mu \nu}$ are,
\begin{align}
    \bigl(R_{1,\mathcal{\tilde B}_{0} + \mathcal{\tilde C}_{0}}^{\mu \nu}\bigr)_{(p_1 \wedge \bar{q})}=&\, Q_1^2Q_2^2(p_1 \cdot p_2)^2 \int d\mu_{\bar{q}} \ \ \bar{q}^2 \ e^{-i\bar{q}.b} (p_1 \wedge \bar{q})^{\mu \nu} \nonumber \int \frac{\hat{d}^4\bar{l}}{\bar{l}^2_+ ((\bar{q}-\bar{l})^2_+)^2} \nonumber\\
    & \times  \left[\frac{m_2^2}{D}\left(-\frac{1}{(p_1\cdot\bar l)_- (p_2\cdot\bar l)_-}+\frac{1}{(p_1\cdot\bar l)_- (p_2\cdot\bar l)_+}\right)\right.\nonumber\\
    &+\frac{(p_1 \cdot p_2)}{D}\left(-\frac{1}{(p_1\cdot\bar l)_- (p_2\cdot\bar l)_-}+\frac{1}{(p_1\cdot\bar l)_- (p_2\cdot\bar l)_+}\right) \nonumber\\
    &\left.+\frac{(\bar q-\bar l)^2 }{(p_1\cdot\bar l)_-^3}\left(\frac{1}{(p_2\cdot\bar l)_-}-\frac{1}{(p_2\cdot\bar l)_+}\right)+\frac{(\bar q-\bar l)^2 }{(p_1\cdot\bar l)_-^2 (p_2\cdot\bar l)_+^2}\right].\label{b.10}
\end{align}
$R_2^{\mu \nu}$ contains only $(p_1 \wedge p_2)^{\mu \nu}$ and $(p_1 \wedge \bar{q})^{\mu \nu}$ terms. First, we write the terms with $(p_1 \wedge p_2)^{\mu \nu}$.

\begin{align}
    \bigl(R_{2,\mathcal{\tilde B}_1 + \mathcal{\tilde C}_1}^{\mu \nu}\bigr)_{p_1 \wedge p_2} =&\, \frac{iQ_1^{2}Q_2^{2}}{2} \frac{(p_1.p_2)^3}{D} (p_1 \wedge p_2)^{\mu \nu} \int d \mu_{\bar{q}} \ e^{-i\bar{q}.b} \int \frac{\hat{d}^4\bar{l}}{\bar{l}^2_+(\bar{q}-\bar{l})^2_+} \nonumber \\
    & \times \biggl[\frac{\bar{l}^2}{(p_1.\bar{l})^2_+} \hat{\delta}(p_2.\bar{l}) + \frac{(\bar{q}-\bar{l})^2}{(p_1.\bar{l})^2_-} \hat{\delta}(p_2.\bar{l}) + \frac{\bar{l}^2}{(p_2.\bar{l})^2_-} \hat{\delta}(p_1.\bar{l}) + \frac{(\bar{q}-\bar{l})^2}{(p_2.\bar{l})^2_+} \hat{\delta}(p_1.\bar{l}) \biggr].\label{b.11}
\end{align}
Now, considering terms containing $(p_1 \wedge \bar{q})^{\mu \nu}$, we have the following,
\begin{align}
    \bigl(R_{2,\mathcal{\tilde B}_1 + \mathcal{\tilde C}_1}^{\mu \nu}\bigr)_{(p_1 \wedge \bar{q})}=&\, Q_1^2Q_2^2 (p_1.p_2) \int d \mu_{\bar{q}} \ e^{-i\bar{q}.b} (p_1 \wedge \bar{q})^{\mu \nu} \int \hat{d}^4\bar{l} \nonumber \\ 
    & \times \left[\frac{(p_1 \cdot p_2) m_2^2}{D}\frac{1}{((\bar q-\bar l)^2_+)^2(p_1\cdot \bar l)_+ }\left(-\frac{1}{(p_2\cdot \bar l)_-}+\frac{1}{(p_2\cdot \bar l)_+}\right) \right. \nonumber\\
    &+\frac{(p_1 \cdot p_2)}{\bar l^2 (p_1\cdot \bar l)_-^3}\left(\frac{1}{(p_2\cdot \bar l)_-}-\frac{1}{(p_2\cdot \bar l)_+}\right) \nonumber\\
    &+\frac{(p_1 \cdot p_2)^2}{D((\bar q-\bar l)^2_+)^2 (p_2\cdot \bar l)_-}\left(\frac{1}{(p_1\cdot \bar l)_-}-\frac{1}{(p_1\cdot \bar l)_+}\right) \nonumber \\
    &-\frac{3}{2\bar l^2 (\bar q-\bar l)^2_+}\left(\frac{1}{(p_1\cdot \bar l)_- (p_2\cdot \bar l)_-}+\frac{1}{(p_1\cdot \bar l)_+ (p_2\cdot \bar l)_+}\right) \nonumber \\
    &+\frac{1}{2\bar l^2 (\bar q-\bar l)^2_+}\left(\frac{1}{(p_1\cdot \bar l)_+ (p_2\cdot \bar l)_-}+\frac{1}{(p_1\cdot \bar l)_- (p_2\cdot \bar l)_+}\right) \nonumber \\
    &+\left.\frac{(p_1 \cdot p_2)}{2 \bar l^2 (p_1\cdot \bar l)_-^2 (p_2\cdot \bar l)_+^2}+\frac{(p_1 \cdot p_2)}{2 (\bar q-\bar l)^2_+ (p_1\cdot \bar l)_-^2 (p_2\cdot \bar l)_-^2}\right].\label{b.12}
\end{align}
Note that, the log divergent pieces in \ref{b.6} and \ref{b.12} cancel among each other,
\begin{align}
    \bigl(R_{1,\mathcal{\tilde B}_1 + \mathcal{\tilde C}_1}^{\mu \nu}\bigr)_{(p_1 \wedge \bar{q})} \Big|_{\text{ln}} + \bigl(R_{2,\mathcal{B}_1 + \mathcal{C}_1}^{\mu \nu}\bigr)_{(p_1 \wedge \bar{q})}\Big|_{\text{ln}}  =&\, 0.\label{b.13}
\end{align}
Finally, the terms containing $(p_1 \wedge p_2)^{\mu \nu}$ and $(p_1 \wedge \bar{q})^{\mu \nu}$ in $R_{2,\mathcal{\tilde B}_{0} + \mathcal{\tilde C}_{0}}^{\mu \nu}$ are as follows,
\begin{align}
    \bigl(R_{2,\mathcal{\tilde B}_{0} + \mathcal{\tilde C}_{0}}^{\mu \nu}\bigr)_{(p_1 \wedge p_2)} =&\, -\frac{Q_1^2Q_2^2}{2} \frac{(p_1.p_2)^3}{D} (p_1 \wedge p_2)^{\mu \nu} \int d \mu_{\bar{q}} \ \bar{q}^2 \ e^{-i\bar{q}.b} \int \frac{\hat{d}^4\bar{l}}{\bar{l}^2_+ (\bar{q}-\bar{l})^2_+} \nonumber \\ 
    & \times \biggl(\frac{i \hat{\delta}(p_2.\bar{l})}{(p_1.\bar{l})^2_+} \biggr) - \frac{Q_1^2Q_2^2}{2} \biggl(\frac{(p_1.p_2)^3+(p_1.p_2)^2 m_1^2}{D}\biggr) (p_1 \wedge p_2)^{\mu \nu} \int d \mu_{\bar{q}} \ \bar{q}^2 \ e^{-i\bar{q}.b} \nonumber \\
    & \times  \int \frac{\hat{d}^4\bar{l}}{\bar{l}^2_+ (\bar{q}-\bar{l})^2_+} \biggl(\frac{i \hat{\delta}(p_2.\bar{l})}{(p_1.\bar{l})^2_-} \biggr).\label{b.14}
\end{align}
And the terms with $(p_1 \wedge \bar{q})^{\mu \nu}$ is
\begin{align}
    \bigl(R_{2,\mathcal{\tilde B}_{0} + \mathcal{\tilde C}_{0}}^{\mu \nu}\bigr)_{(p_1 \wedge \bar{q})} =&\, Q_1^2Q_2^2 \int d \mu_{\bar{q}} \ e^{-i\bar{q}.b} (p_1 \wedge \bar{q})^{\mu \nu} \int \frac{\hat{d}^4\bar{l}}{\bar{l}^2_+ (\bar{q}-\bar{l})^2_+} \nonumber \\ 
    & \times \left[ \frac{(p_1 \cdot p_2)^2 m_2^2 \bar q^2}{D (\bar q-\bar l)^2_+} \Bigg(\frac{1}{(p_1\cdot\bar l)_- (p_2\cdot\bar l)_-}+\frac{1}{(p_1\cdot\bar l)_+ (p_2\cdot\bar l)_-}-\frac{1}{(p_1\cdot\bar l)_- (p_2\cdot\bar l)_+} \right. \nonumber \\
    & -\frac{1}{(p_1\cdot\bar l)_+ (p_2\cdot\bar l)_+} \Bigg)+\frac{(p_1 \cdot p_2)^2 m_2^2}{D} \Bigg(\frac{1}{(p_1\cdot\bar l)_- (p_2\cdot\bar l)_-} -\frac{1}{(p_1\cdot\bar l)_+ (p_2\cdot\bar l)_-} \nonumber \\
    & - \frac{1}{(p_1\cdot\bar l)_- (p_2\cdot\bar l)_+} + \frac{1}{(p_1\cdot\bar l)_+ (p_2\cdot\bar l)_+}\Bigg) + \frac{(p_1 \cdot p_2)^3}{D} \Bigg(\frac{1}{(p_1\cdot\bar l)_- (p_2\cdot\bar l)_-} \nonumber \\
    &- \frac{1}{(p_1\cdot\bar l)_+ (p_2\cdot\bar l)_-} - \frac{1}{(p_1\cdot\bar l)_- (p_2\cdot\bar l)_+} + \frac{1}{(p_1\cdot\bar l)_+ (p_2\cdot\bar l)_+}\Bigg)\nonumber \\
    &+ \frac{(p_1 \cdot p_2)^3 \bar q^2}{D (\bar q-\bar l)^2_+} \Bigg(\frac{1}{(p_1\cdot\bar l)_+ (p_2\cdot\bar l)_-}-\frac{1}{(p_1\cdot\bar l)_- (p_2\cdot\bar l)_+}\Bigg) \nonumber\\
    &\left. + \frac{(p_1 \cdot p_2)^2 \bar q^2}{(p_1\cdot\bar l)_-^3}\left(-\frac{1}{(p_2\cdot\bar l)_-}+\frac{1}{(p_2\cdot\bar l)_+}\right)-\frac{(p_1 \cdot p_2)^2 \bar q^2}{2 (p_1\cdot\bar l)_-^2 (p_2\cdot\bar l)_+^2} \right] .\label{b.15}
\end{align}

\section{$C^{\mu \nu}$ at one loop}\label{D}
Here we explicitly evaluate terms contained in the cut contribution, beginning with $C_{11}^{\mu \nu}$
\begin{align}
    C_{11}^{\mu \nu} =&\, \frac{i Q_1^2Q_2^2 \ \hbar^{-1}}{16} \int \hat{d}^4\bar{q} \ \hat{d}^4\bar{w}\, e^{-i\bar{q}.b}\ \biggl[\hat\delta(p_1\cdot \bar q)\hat\delta(p_2\cdot \bar q)\hat\delta(p_1\cdot \bar w)\hat\delta(p_2\cdot \bar w) \nonumber\\
    &+ \hbar \left(\hat\delta(p_1\cdot \bar q)\hat\delta(p_2\cdot \bar q)\left(\hat\delta'(p_1\cdot \bar w)\hat\delta(p_2\cdot \bar w)-\hat\delta(p_1\cdot \bar w)\hat\delta'(p_2\cdot \bar w)\right)\right. \nonumber \\ 
    &+ \left.\left(\hat\delta'(p_1\cdot \bar q)\hat\delta(p_2\cdot \bar q)-\hat\delta(p_1\cdot \bar q)\hat\delta'(p_2\cdot \bar q)\right)\hat\delta(p_1\cdot \bar w)\hat\delta(p_2\cdot \bar w)\right) \biggr] \nonumber\\
    &\times\,  \biggl(\frac{4p_1.p_2 + 2 \hbar (p_2-p_1).(\bar{w} + \bar{q})}{(\bar{q} - \bar{w})^2_+}\biggr) \biggl(\frac{4(p_1 \wedge p_2)^{\mu \nu} - 2 \hbar (p_1 \wedge \bar{w})^{\mu \nu}}{\bar{w}^2_+}\biggr).\label{c.1}
\end{align}
The classical term here is,
\begin{align}
    C_{11}^{\mu \nu} =&\, iQ_1^2Q_2^2 \ (p_1.p_2) \ (p_1 \wedge p_2)^{\mu \nu} \int \frac{\hat{d}^4\bar{q} \ \hat{d}^4\bar{w}}{\bar{w}^2_+ (\bar{q}-\bar{w})^2_+}e^{-i\bar{q}.b} \nonumber\\
    &\times\biggl[\hat\delta(p_1\cdot \bar q)\hat\delta(p_2\cdot \bar q)\left(\hat\delta'(p_1\cdot \bar w)\hat\delta(p_2\cdot \bar w)-\hat\delta(p_1\cdot \bar w)\hat\delta'(p_2\cdot \bar w)\right) \nonumber \\ 
    &+ \left(\hat\delta'(p_1\cdot \bar q)\hat\delta(p_2\cdot \bar q)-\hat\delta(p_1\cdot \bar q)\hat\delta'(p_2\cdot \bar q)\right)\hat\delta(p_1\cdot \bar w)\hat\delta(p_2\cdot \bar w)\biggr] \nonumber\\
    &- \frac{iQ_1^2Q_1^2}{2} \ (p_1.p_2) \int \frac{d\mu_{\bar q,\bar w}}{\bar{w}^2_+ (\bar{q}-\bar{w})^2_+} \,e^{-i\bar{q}.b}\, (p_1 \wedge \bar{w})^{\mu \nu} .\label{c.2}
\end{align}
The first term here vanishes. Now, in this integral, we do the following transformation $\bar{w} \rightarrow \bar{q}-\bar{w}$ which exchanges the photon momenta between the two legs and then average over them. After averaging, we have the following,
\begin{align}
    C_{11}^{\mu \nu} =&\, - \frac{iQ_1^2Q_1^2}{4} \ (p_1.p_2) \int \frac{d\mu_{\bar q,\bar w}}{\bar{w}^2_+ (\bar{q}-\bar{w})^2_+} \,e^{-i\bar{q}.b}\, (p_1 \wedge \bar{q})^{\mu \nu}  \nonumber \\
    =&\, \frac{-Q_1Q_2}{8\pi \sqrt{D}} (p_1 \wedge \Delta p)^{\mu \nu} \ \log\omega \ .\label{c.3}
\end{align}
The term $C_{12}^{\mu \nu}$ has the following expression,
\begin{align}
    C_{12}^{\mu \nu} =&\, \frac{-iQ_1^2Q_2^2}{8} \int \hat{d}^4\bar{q} \ \hat{d}^4\bar{w} \ e^{-i\bar{q}.b} \biggl[\hat\delta(p_1\cdot \bar q)\hat\delta(p_2\cdot \bar q)\hat\delta(p_1\cdot \bar w)\hat\delta(p_2\cdot \bar w) \nonumber\\
    &+ \hbar \bigl(\hat\delta(p_1\cdot \bar q)\hat\delta(p_2\cdot \bar q)\left(\hat\delta'(p_1\cdot \bar w)\hat\delta(p_2\cdot \bar w)-\hat\delta(p_1\cdot \bar w)\hat\delta'(p_2\cdot \bar w)\right) \nonumber \\
    &+ \left(\hat\delta'(p_1\cdot \bar q)\hat\delta(p_2\cdot \bar q)-\hat\delta(p_1\cdot \bar q)\hat\delta'(p_2\cdot \bar q)\right)\hat\delta(p_1\cdot \bar w)\hat\delta(p_2\cdot \bar w)\bigr) \biggr]\nonumber\\
    &\times\,\frac{\bar{w} \wedge p_2}{(\bar{q}-\bar{w})^2_+} \frac{4p_1.p_2 + 2\hbar(p_2 - p_1).\bar{w}}{\bar{w}^2_+}.\label{c.4}
\end{align}
The classical term that survives,
\begin{align}
    C_{12}^{\mu \nu} =&\, \frac{iQ_1^2Q_2^2}{2} (p_1.p_2) \int \frac{d\mu_{\bar q,\bar w}}{\bar{w}^2_+ (\bar{q}-\bar{w})^2_+} \,e^{-i\bar{q}.b}\, (p_2 \wedge \bar{w})^{\mu \nu}.\label{c.5}
\end{align}
Similarly, for this integral, after averaging over the two, we have the following,
\begin{align}
    C_{12}^{\mu \nu} =&\, \frac{iQ_1^2Q_2^2}{4} (p_1.p_2) \int \frac{d\mu_{\bar q,\bar w}}{\bar{w}^2_+ (\bar{q}-\bar{w})^2_+} \,e^{-i\bar{q}.b}\, (p_2 \wedge \bar{q})^{\mu \nu} \nonumber \\
    =&\, \frac{Q_1Q_2}{8 \pi \sqrt{D}} (p_2 \wedge \Delta p)^{\mu \nu} \ \log\omega.\label{c.6}
\end{align}
Next we move onto calculating $C_2^{\mu \nu}$, expanding $C_{21}^{\mu \nu}$ in $\hbar$ we have the following,
\begin{align}
    C_{21}^{\mu \nu} =&\, \frac{iQ_1^2Q_2^2 \ \hbar^{-1}}{16} \int \hat{d}^4\bar{q} \ \hat{d}^4\bar{w} \biggl\{\hat{\delta}'(p_1.\bar{q}) \hat{\delta}(p_1.\bar{w}) + \hat{\delta}(p_1.\bar{q}) \hat{\delta}'(p_1.\bar{w}) + \frac{\hbar}{2}\biggl(\bar{w}^2 \bigl( \hat{\delta}'(p_1.\bar{q}) \hat{\delta}'(p_1.\bar{w}) \nonumber \\
    &+ \hat{\delta}(p_1.\bar{q}) \hat{\delta}''(p_1.\bar{w}) \bigr) + \bar{q}^2 \bigl(\hat{\delta}''(p_1.\bar{q}) \hat{\delta}(p_1.\bar{w}) + \hat{\delta}'(p_1.\bar{q}) \hat{\delta}'(p_1.\bar{w})\bigr) \biggr)\biggr\} \biggl\{\hat{\delta}(p_2.\bar{q}) \hat{\delta}(p_2.\bar{w}) \nonumber \\
    &- \frac{\hbar}{2} \biggl(\bar{w}^2 \hat{\delta}(p_2.\bar{q}) \hat{\delta}'(p_2.\bar{w}) + \bar{q}^2  \hat{\delta}'(p_2.\bar{q}) \hat{\delta}(p_2.\bar{w})\biggr) \biggr\} (p_1 \wedge \bar{w})^{\mu \nu} \ e^{-i\bar{q}.b} \biggl(\frac{4p_1.p_2}{\bar{w}^2_+} \nonumber \\
    &+ \frac{2\hbar (p_2 - p_1).\bar{w}}{\bar{w}^2_+} \biggr) \biggl(\frac{4p_1.p_2 + 2\hbar (p_2 - p_1).(\bar{w}+\bar{q})}{(\bar{q}-\bar{w})^2_+}\biggr).\label{c.8}
\end{align}
The classical components that survive from $C_{21}^{\mu \nu}$ are the following,
\begin{align}
    C_{21}^{\mu \nu} =&\, -iQ_1^2Q_2^2 (p_1.p_2) \int \frac{\hat{d}^4\bar{q} \ \hat{d}^4\bar{w}}{\bar{w}^2_+ (\bar{q}-\bar{w})^2_+} e^{-i\bar{q}.b} \ (p_1 \wedge \bar{w})^{\mu \nu} \hat{\delta}(p_1.\bar{q})  \hat{\delta}'(p_1.\bar{w}) \hat{\delta}(p_2.\bar{q}) \hat{\delta}(p_2.\bar{w}) \bigl(p_1.\bar{w}\bigr) \nonumber \\
    & - \frac{iQ_1^2Q_2^2}{2} (p_1.p_2) \int \frac{\hat{d}^4\bar{q} \ \hat{d}^4\bar{w}}{\bar{w}^2_+ (\bar{q}-\bar{w})^2_+} e^{-i\bar{q}.b} \ (p_1 \wedge \bar{w})^{\mu \nu} \hat{\delta}'(p_1.\bar{q}) \hat{\delta}(p_1.\bar{w}) \hat{\delta}(p_2.\bar{q}) \hat{\delta}(p_2.\bar{w}) \bigl(p_1. \bar{q}\bigr) \nonumber \\
    & + \frac{iQ_1^2Q_2^2}{2} (p_1.p_2)^2 \int \frac{\hat{d}^4\bar{q} \ \hat{d}^4\bar{w}}{(\bar{q}-\bar{w})^2_+} e^{-i\bar{q}.b} \ (p_1 \wedge \bar{w})^{\mu \nu} \biggl(\hat{\delta}(p_1.\bar{q})  \hat{\delta}''(p_1.\bar{w}) \hat{\delta}(p_2.\bar{q}) \hat{\delta}(p_2.\bar{w}) \nonumber \\
    & + \hat{\delta}'(p_1.\bar{q}) \hat{\delta}'(p_1.\bar{w}) \hat{\delta}(p_2.\bar{q}) \hat{\delta}(p_2.\bar{w}) - \hat{\delta}'(p_1.\bar{q})  \hat{\delta}(p_1.\bar{w}) \hat{\delta}(p_2.\bar{q}) \hat{\delta}'(p_2.\bar{w}) \nonumber \\
    &- \hat{\delta}(p_1.\bar{q}) \hat{\delta}'(p_1.\bar{w}) \hat{\delta}(p_2.\bar{q}) \hat{\delta}'(p_2.\bar{w})\biggr) + \frac{iQ_1^2Q_2^2}{2} (p_1.p_2)^2 \int \frac{\hat{d}^4\bar{q} \ \hat{d}^4\bar{w}}{\bar{w}^2_+ (\bar{q}-\bar{w})^2_+} \bar{q}^2 e^{-i\bar{q}.b} \ (p_1 \wedge \bar{w})^{\mu \nu} \nonumber \\
    &\times \biggl(\hat{\delta}'(p_1.\bar{q}) \hat{\delta}'(p_1.\bar{w}) \hat{\delta}(p_2.\bar{q}) \hat{\delta}(p_2.\bar{w})  + \hat{\delta}''(p_1.\bar{q}) \hat{\delta}(p_1.\bar{w}) \hat{\delta}(p_2.\bar{q}) \hat{\delta}(p_2.\bar{w}) \nonumber \\
    & - \hat{\delta}(p_1.\bar{q}) \hat{\delta}'(p_1.\bar{w})  \hat{\delta}'(p_2.\bar{q}) \hat{\delta}(p_2.\bar{w}) - \hat{\delta}'(p_1.\bar{q}) \hat{\delta}'(p_1.\bar{w}) \hat{\delta}'(p_2.\bar{q}) \hat{\delta}(p_2.\bar{w})\biggr).\label{c.9}
\end{align}
After doing IBP none of the terms containing $(p_1 \wedge p_2)^{\mu \nu}$ survive from here, the remaining terms contain only $(p_1 \wedge \bar{w})^{\mu \nu}$, so
\begin{align}
    \biggl(C_{21}^{\mu \nu}\biggr)_{p_1 \wedge \bar{w}} =&\, \frac{3iQ_1^2Q_2^2}{2} (p_1.p_2) \int \frac{d\mu_{\bar q,\bar w}}{\bar{w}^2_+ (\bar{q}-\bar{w})^2_+} e^{-i\bar{q}.b} \ (p_1 \wedge \bar{w})^{\mu \nu} \nonumber \\
    & - iQ_1^2Q_2^2 \biggl(\frac{m_2^2(p_1.p_2)^2 + (p_1.p_2)^3}{D}\biggr) \int \frac{d\mu_{\bar q,\bar w}}{\bar{w}^2_+ (\bar{q}-\bar{w})^2_+} e^{-i\bar{q}.b}\,(p_1 \wedge \bar{w})^{\mu \nu} \nonumber \\
    &  -iQ_1^2Q_2^2 \biggl(\frac{m_2^2(p_1.p_2)^2 + (p_1.p_2)^3}{D}\biggr)  \int \frac{d\mu_{\bar q,\bar w}}{((\bar{q}-\bar{w})^2_+)^2} e^{-i\bar{q}.b} \ (p_1 \wedge \bar{w})^{\mu \nu} .\label{c.10}
\end{align}
Now the term $C_{22}^{\mu \nu}$ has only one classical term which is,
\begin{align}
    C_{22}^{\mu \nu} =&\, iQ_1^2Q_2^2(p_1.p_2)^2 \int \frac{\hat{d}^4\bar{q} \ \hat{d}^4\bar{w}}{\bar{w}^2_+ (\bar{q}-\bar{w})^2_+} e^{-i\bar{q}.b} \ (\bar{q} \wedge \bar{w})^{\mu \nu} \hat{\delta}'(p_1.\bar{q}) \hat{\delta}(p_1.\bar{w}) \hat{\delta}(p_2.\bar{q}) \hat{\delta}(p_2.\bar{w}).\label{c.11}
\end{align}
There is no $(p_1 \wedge p_2)^{\mu \nu}$ term here so,
\begin{align}
    \biggl(C_{22}^{\mu \nu}\biggr)_{p_1 \wedge \bar{w}} =&\, iQ_1^2Q_2^2 \frac{m_2^2(p_1.p_2)^2}{D} \int \frac{d\mu_{\bar q,\bar w}}{\bar{w}^2_+ (\bar{q}-\bar{w})^2_+} e^{-i\bar{q}.b} \ (p_1 \wedge \bar{w})^{\mu \nu}  \nonumber \\
    & - iQ_1^2Q_2^2 \frac{(p_1.p_2)^3}{D} \int \frac{d\mu_{\bar q,\bar w}}{\bar{w}^2_+ (\bar{q}-\bar{w})^2_+} e^{-i\bar{q}.b} \ (p_2 \wedge \bar{w})^{\mu \nu} .\label{c.12}
\end{align}
Finally, $C_{23}^{\mu \nu}$ has the following form,
\begin{align}
    C_{23}^{\mu \nu} =&\, \frac{-iQ_1^2Q_2^2}{8} \int \hat{d}^4\bar{q} \ \hat{d}^4\bar{w} \ \biggl(\hat\delta(p_1\cdot \bar q)\hat\delta(p_2\cdot \bar q) \ + \ \frac{\hbar}{2} \left(\hat\delta'(p_1\cdot \bar q)\hat\delta(p_2\cdot \bar q)-\hat\delta(p_1\cdot \bar q)\hat\delta'(p_2\cdot \bar q)\right)\biggr) \nonumber\\
    &\times\biggl(\hat\delta(p_1\cdot \bar w)\hat\delta(p_2\cdot \bar w) \ + \ \frac{\hbar}{2} \left(\hat\delta'(p_1\cdot \bar w)\hat\delta(p_2\cdot \bar w)-\hat\delta(p_1\cdot \bar w)\hat\delta'(p_2\cdot \bar w)\right)\biggr) e^{-i\bar{q}.b} \nonumber \\
    &\times \frac{(\bar{w} \wedge p_2)^{\mu \nu}}{(\bar{w} - \bar{q})^2_+} \biggl(\frac{4p_1.p_2 + 2\hbar(p_2 - p_1).\bar{w}}{\bar{w}^2_+}\biggr).\label{c.13}
\end{align}
The only classical term from this part,
\begin{align}
    \biggl(C_{23}^{\mu \nu}\biggr)_{p_1 \wedge \bar{w}} =&\, \frac{iQ_1^2Q_2^2}{2} (p_1.p_2) \int \frac{d\mu_{\bar q,\bar w}}{\bar{w}^2_+ (\bar{q}-\bar{w})^2_+} e^{-i\bar{q}.b} \ (p_2 \wedge \bar{w})^{\mu \nu}.\label{c.14}
\end{align}
The terms containing $\frac{1}{((\bar{q} - \bar{w})^2_+)^2}$ in $C_{21}^{\mu \nu}$ do not contribute as they are not log divergent, so the final result is,
\begin{align}
    C_2^{\mu \nu} =&\, \frac{3iQ_1^2Q_2^2}{2} (p_1.p_2) \int \frac{d\mu_{\bar q,\bar w}}{\bar{w}^2_+ (\bar{q}-\bar{w})^2_+} e^{-i\bar{q}.b} \ (p_1 \wedge \bar{w})^{\mu \nu} \nonumber \\
    & + \frac{iQ_1^2Q_2^2}{2} (p_1.p_2) \int \frac{d\mu_{\bar q,\bar w}}{\bar{w}^2_+ (\bar{q}-\bar{w})^2_+} e^{-i\bar{q}.b} \ (p_2 \wedge \bar{w})^{\mu \nu}  \nonumber \\
    & - iQ_1^2Q_2^2 \frac{(p_1.p_2)^3}{D} \int \frac{d\mu_{\bar q,\bar w}}{\bar{w}^2_+ (\bar{q}-\bar{w})^2_+} e^{-i\bar{q}.b} \ (p_2 \wedge \bar{w})^{\mu \nu}  \nonumber \\
    & - iQ_1^2Q_2^2 \frac{(p_1.p_2)^3}{D} \int \frac{d\mu_{\bar q,\bar w}}{\bar{w}^2_+ (\bar{q}-\bar{w})^2_+} e^{-i\bar{q}.b} \ (p_1 \wedge \bar{w})^{\mu \nu} .\label{c.15}
\end{align}
Finally, averaging over the transformation $\bar{w} \rightarrow \bar{q} - \bar{w}$ we have,
\begin{align}
    C_2^{\mu \nu} =&\, \frac{3iQ_1^2Q_2^2}{4} (p_1.p_2) \int \frac{\hat{d}^4\bar{q} \ \hat{d}^4\bar{w}}{\bar{w}^2_+ (\bar{q}-\bar{w})^2_+} e^{-i\bar{q}.b} \ (p_1 \wedge \bar{q})^{\mu \nu}  \nonumber \\
    & + \frac{iQ_1^2Q_2^2}{4} (p_1.p_2) \int \frac{\hat{d}^4\bar{q} \ \hat{d}^4\bar{w}}{\bar{w}^2_+ (\bar{q}-\bar{w})^2_+} e^{-i\bar{q}.b} \ (p_2 \wedge \bar{q})^{\mu \nu}  \nonumber \\
    & - iQ_1^2Q_2^2 \frac{(p_1.p_2)^3}{2D} \int \frac{\hat{d}^4\bar{q} \ \hat{d}^4\bar{w}}{\bar{w}^2_+ (\bar{q}-\bar{w})^2_+} e^{-i\bar{q}.b} \ (p_2 \wedge \bar{q})^{\mu \nu}  \nonumber \\
    & - iQ_1^2Q_2^2 \frac{(p_1.p_2)^3}{2D} \int \frac{\hat{d}^4\bar{q} \ \hat{d}^4\bar{w}}{\bar{w}^2_+ (\bar{q}-\bar{w})^2_+} e^{-i\bar{q}.b} \ (p_1 \wedge \bar{q})^{\mu \nu}  \nonumber \\
    =&\, \biggl[ \frac{Q_1Q_2}{8 \pi \sqrt{D}} \bigl((3p_1 + p_2) \wedge \Delta p \bigr)^{\mu \nu} - Q_1Q_2 \frac{(p_1 . p_2)^2}{4\pi D^{\frac{3}{2}}} \bigl((p_1+p_2) \wedge \Delta p \bigr)^{\mu \nu} \biggr] \ \log\omega.\label{c.16}
\end{align}

%%%%%%%%%%%%%%%%%%%%%%%%%%%%%%%%%%%%%%%%%%%%%%
\newpage
\bibliographystyle{JHEP}
\bibliography{NLOconstraint}

\end{document}